%% file: main.tex
\documentclass[conference, compsoc]{IEEEtran}
\usepackage{graphicx}   
\usepackage{caption}
\usepackage{subcaption} 

\usepackage{comment}
\usepackage{amssymb}  
\newtheorem{theorem}{Theorem}

\usepackage{flushend}
\usepackage{mathtools}
\usepackage{aligned-overset}

\usepackage{pifont}
\newcommand{\xmark}{\ding{55}}

\usepackage{algpseudocode}
\usepackage{amsmath}

\usepackage{wasysym}

\allowdisplaybreaks
\usepackage{lipsum}

\usepackage[table]{xcolor}
\definecolor{lightblue}{RGB}{220, 230, 241}
\definecolor{lightyellow}{RGB}{255, 255, 204}

\usepackage{enumitem,kantlipsum}
\usepackage{multirow,hhline}
\usepackage{colortbl}
\usepackage{listings}
\usepackage{makecell}
\usepackage{hyperref}
\IEEEoverridecommandlockouts
\usepackage{xspace}
\hypersetup{colorlinks=true,
            linkcolor=black,
            anchorcolor=black,
            citecolor=black,
            urlcolor=black
            }
\usepackage{url}

\usepackage{cuted}
\usepackage{cleveref}
\allowdisplaybreaks

\usepackage[linesnumbered,ruled,noline]{algorithm2e}

\usepackage{tikz}

\SetCommentSty{mycommfont}
\SetKwInput{KwInput}{Input}                
\SetKwInput{KwOutput}{Output}              
\SetKwBlock{Loop}{Loop}{}
\SetKwBlock{Func}{Function}{}
\SetKwBlock{If}{If}{}

\ifCLASSOPTIONcompsoc
  \usepackage[nocompress]{cite}
\else
  \usepackage{cite}
\fi

\begin{document}

\newcommand{\pname}{RadKey\xspace}

\title{\pname: An LLM-Guided RF Backscatter System for Through-Wall Keystroke Inference}
\author{
Qijun Wang$^{\ast}$,
Chunqi Qian$^{\dagger}$,
and Huacheng Zeng$^{\ast}$\\
$^{\ast}$Department of Computer Science and Engineering, Michigan State University\\
$^{\dagger}$Department of Radiology, Michigan State University
}

\maketitle

\begin{abstract}
\input{abstract}
\end{abstract}

\IEEEpeerreviewmaketitle


\section{Introduction}
\input{1_introduction}

\section{Threat Model}

\input{3_threat_model_attack_overview}

\section{Understanding Keystrokes}

\input{4_preliminary_study}

\section{RF Tag Design}
\input{5_RFtag_based_vibration_acquisition_and_feature_extraction}

\section{RF Reader Design}

\input{6_LLM-powered_privacy_information_inference}

\section{Experimental Evaluation}
\input{7_evaluation}

\section{Countermeasure Strategies}
\input{8_defence_against_keystroking_attack}

\section{Related Work}
\input{2_related_work}

\section{Conclusion}
\input{9_conclusion}

\input{10_appendix}

\section*{Acknowledgments}
The authors sincerely thank the anonymous reviewers and the shepherd for their valuable comments and constructive feedback. 
The work of Q.~Wang and H.~Zeng was supported by the NSF under Grants ECCS-2225337 and ECCS-2434001. 
The work of C.~Qian was supported by the NSF under Grant ECCS-2144138.

\section*{Ethics Considerations}
All user studies in this work were conducted in accordance with established ethical guidelines for research involving human participants. The study protocol was reviewed by the Institutional Review Board (IRB) of the authors' institution and was determined to be exempt from IRB oversight. Participants were informed of the study procedures and provided informed consent prior to participation. Participation was voluntary, and participants could withdraw at any time without penalty. All collected data were anonymized to protect participant privacy. No experiments were conducted on unaware individuals or in unauthorized real-world deployments.

\section*{LLM Usage Considerations}
LLMs are used in this work as part of the proposed methodology. In particular, RadKey employs an LLM during online adaptation to generate pseudo ground-truth labels for semantically coherent typing inputs, which are then used to update the keystroke classifier at runtime. The LLM was run locally on a research workstation, and no participant typing data were sent to external services. All technical claims, system design choices, implementations, and experimental results were developed, verified, and validated by the authors. Because LLM outputs may be imperfect and may vary across models and prompting settings, the LLM is used only as an auxiliary signal in the adaptation loop rather than as definitive ground truth.

\bibliographystyle{IEEEtran}
\bibliography{references}

\appendices

\newpage 

\end{document}

%% file: abstract.tex
In today's digitally connected world, keyboards remain the primary interface for inputting sensitive information, making them a persistent target for eavesdropping attacks. While prior keystroke inference techniques have exploited side-channel signals such as acoustics and vibrations, they typically rely on conspicuous, short-range sensors and require victim-specific data for model training, limiting their practicality, scalability, and stealth.
In this paper, we present \pname, an RF backscatter system for covert, long-range, through-wall keystroke eavesdropping. \pname comprises two components: a compact batteryless backscatter tag and an RF reader. 
The tag captures keystroke-induced vibrations and acoustic signals, modulating them onto the frequency shift of its backscattered RF signal using two magnetically-coupled LC resonators. This design also enables spectral separation between the excitation and backscatter signals, mitigating self-interference for the RF reader and thus extending eavesdropping range.
The RF reader demodulates the backscattered RF signal to infer typed content.
It employs a dedicated signal processing pipeline that extracts user- and keyboard-independent keystroke features across time and frequency domains, enabling strong generalizability. To further enhance adaptability, \pname integrates an LLM for online adaptation, leveraging LLM outputs as pseudo ground-truth labels to refine the classifier during runtime.
We have built a prototype of the full \pname system and evaluated it through extensive over-the-air experiments. Results show that \pname achieves accurate and robust keystroke inference across diverse users in real-world settings.
A demo video is available at:
\url{https://radkey-submission.github.io/RadKey/}

%% file: 1_introduction.tex
In our digitally interconnected world, the keyboard remains the primary interface for inputting our most sensitive information, including passwords, financial details, confidential communications, and proprietary data.
\vspace{0.35em}
\noindent\rule{\columnwidth}{0.3pt}
\vspace{0.2em}

{\scriptsize
\noindent
\copyright~2026 IEEE. Personal use of this material is permitted.
Permission from IEEE must be obtained for all other uses, in any current
or future media, including reprinting/republishing this material for
advertising or promotional purposes, creating new collective works, for
resale or redistribution to servers or lists, or reuse of any copyrighted
component of this work in other works.

\vspace{0.2em}
\noindent
Accepted to the 47th IEEE Symposium on Security and Privacy (IEEE S\&P),
2026. DOI: 10.1109/SP63933.2026.00160.
}

\noindent
Consequently, keystroke privacy has emerged as a critical aspect of cybersecurity, with eavesdropping attacks presenting a persistent and evolving threat.
Traditionally, adversaries exploit various side-channel emanations (e.g., acoustic signals \cite{harrison2023practical, feng2025keyprint, luo2024eavesdropping, wang2024reflexnoop, taheritajar2024survey, wang2025batteryless, wang2026radear}, electromagnetic radiation \cite{vuagnoux2009compromising, jiang2024ghosttype, jin2021periscope, zhan2022graphics}, radar-based sensing \cite{zhang2025radsee, zhang2025radeye}, and mechanical vibrations generated during keystroke events \cite{owusu2012accessory, cai2011touchlogger, liu2015good, liu2017vibsense}) to infer user inputs. 
However, despite recent advancements, existing approaches suffer from some fundamental limitations that restrict their practicality for covert and long-term keystroke surveillance.

\begin{figure}
    \centering
    \includegraphics[width=0.999\linewidth]{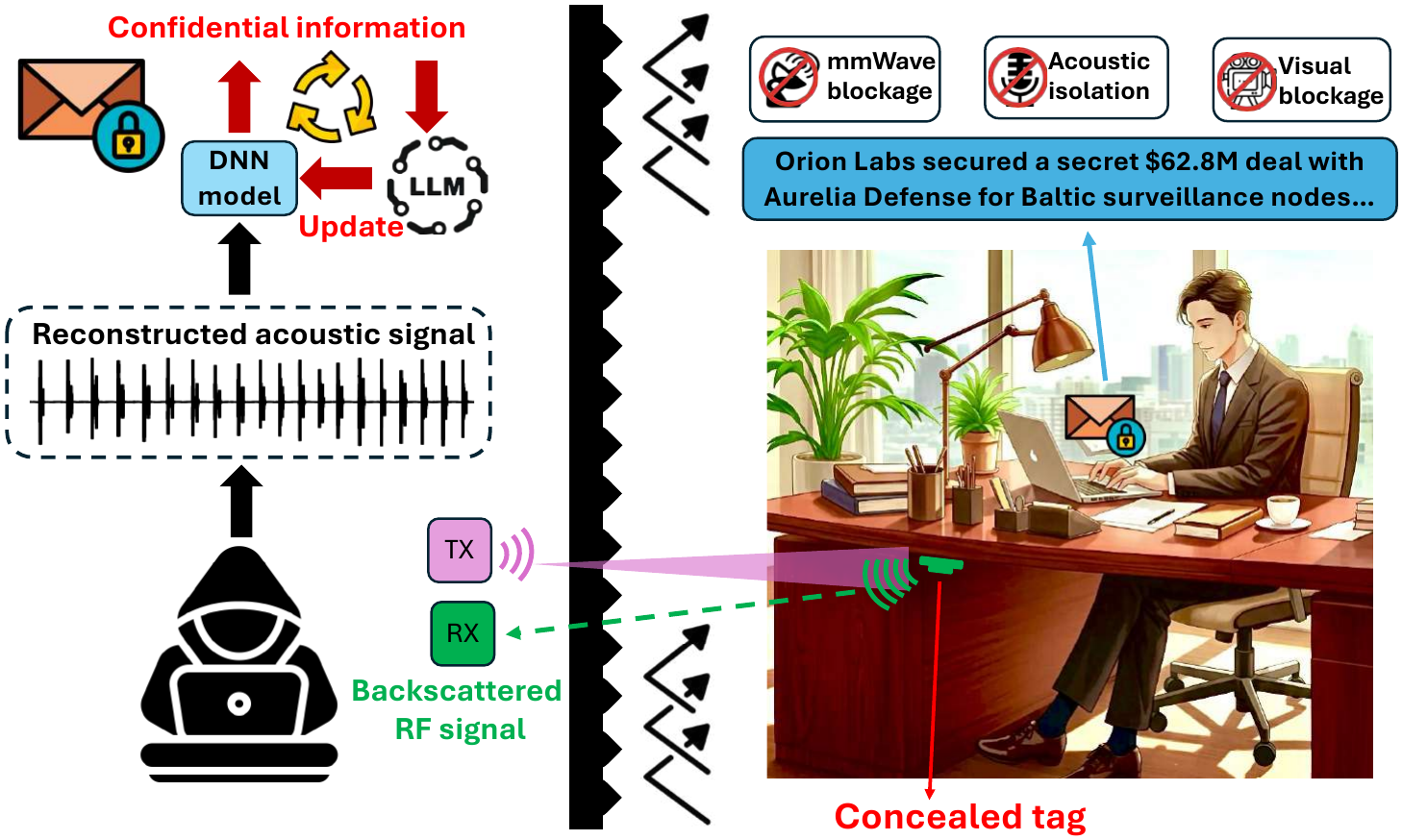}
    \caption{Threat model and system configuration.}
    \label{fig:sys_scenario}
\end{figure}

Most existing keystroke eavesdropping attacks depend on a strategic placement of conspicuous sensing devices (e.g., microphones or smartphones) in close proximity to the target keyboard \cite{huang2024keystrokesniffer, liu2015snooping, bai2021know, lu2019keylistener, chen2022behavicker, wijewickrama2023overhear, zhao2023ultrasnoop, zhu2014context, wang2014ubiquitous}.
Such setups are not only easily discoverable to victims but also impractical for continuous surveillance due to their dependency on batteries or regular maintenance.
Moreover, the effectiveness of attacks typically hinges on extensive user-specific training data collection, requiring attackers to gather large sets of labeled keystroke data from the specific victim and keyboard \cite{ali2015keystroke, meng2019revealing, huang2024keystrokesniffer, wang2014ubiquitous, yang2023towards, maiti2018side, sabra2020zoom, taheritajar2024acoustic}.
This requirement fundamentally restricts the attack's feasibility, adaptability, and scalability.
Some attacks additionally suffer from closed-vocabulary constraints (e.g., \cite{feng2025keyprint, marquardt2011sp, chen2015tracking, huang2024keystrokesniffer, monaco2018sok}), where high accuracy is only achieved under the assumption that user input belongs to a small predefined word set, limiting their generalizability in realistic scenarios.




In this paper, we consider a threat model as illustrated in Fig.~\ref{fig:sys_scenario}, where an adversary seeks to eavesdrop on keystrokes generated by a victim typing on a physical keyboard or a laptop. Based on the captured keystroke signals, the adversary aims to reconstruct the typed content and infer sensitive information such as messages, search queries, or passwords. 
To perform this attack, we propose \pname, an RF backscatter system composed of two key components:
\textit{a passive RF backscatter tag} and \textit{an RF reader}.
The tag is covertly placed within the target environment, mounted underneath the table or desk supporting the keyboard.
It captures keystroke-induced vibrations and modulates them onto a backscattered RF signal, which is then received by the RF reader.
The RF reader, positioned either inside or outside the target space, performs signal processing to demodulate the keystroke-related signal and infer the typed content.
In the design of \pname, we face two challenges.

\textbf{Challenge \#1: RF Tag Design.}
To enable reliable eavesdropping, the RF backscatter tag has the following requirements.
\textit{First}, it must be compact (e.g., under one inch) and easily concealable.
\textit{Second}, it must support \textit{continuous} {keystroke eavesdropping} using only harvested energy. While prior batteryless tags (e.g., WISP~\cite{menon2022wireless}, Battery-Free Phone~\cite{talla2017battery}, MARS~\cite{arora2021mars}) demonstrate RF communication, they operate intermittently with low duty cycles due to limited energy availability.
\textit{Third}, the tag must reliably modulate RF signals to ensure long-range and through-wall performance. A common limitation in existing systems (e.g., RFID) is self-interference: when the backscattered signal shares the same frequency as the excitation signal, it overwhelms the reader and severely reduces detection performance.


To meet these requirements, we introduce a novel dual-resonator architecture for the RF backscatter tag. 
Our proposed tag comprises four elements: 
an off-the-shelf piezoelectric sensor, 
a voltage sensing resonator (VSR), 
a parametric enhancement resonator (PER), 
and 
a dipole antenna.
The piezoelectric sensor transduces sound and vibration-induced fluctuations into a voltage signal, which changes the capacitance of a diode on the VSR and therefore changes the VSR's resonance frequency.
The design of the VSR establishes an approximate linear relationship between the piezoelectric sensor's output voltage and the VSR's resonance frequency change, achieving a direct frequency modulation of the sound/vibration signal. 
The VSR is magnetically coupled to the PER, which serves as an energy pump and thus amplifies the VSR's resonance frequency for radiation. 
More importantly, the PER introduces the spectral separation of the excitation and backscatter signals via voltage-tunable dual-mode resonance.
The modulated signal is then radiated through the dipole antenna coupled to the PER, significantly enhancing the quality of the backscatter signal and therefore boosting the detection range of the RF reader.


The advantages of this tag design are multifaceted.
First, it achieves a substantial spectral separation between excitation and backscatter signals, mitigating the self-interference issue for the RF reader.
Second, the tag supports frequency modulation by establishing a linear relationship between shifts in its resonance frequency and the amplitude of the keystroke signal. This simplifies keystroke demodulation at the RF reader.
Third, the design is compact enough to fit on a one-inch PCB, enabling easy concealment within the target environment.
Finally, the tag can operate at a low frequency (e.g., 915 MHz), which offers strong wall penetration and favorable propagation characteristics for through-wall keystroke detection.

\textbf{Challenge \#2: RF Reader Design.}
Designing an effective RF reader poses several challenges.
First, the backscattered signal is inherently weak due to the passive nature of the tag and the lack of a line-of-sight (LoS) path between the tag and the reader. 
Second, the keystroke classification task involves a large vocabulary. Unlike small-scale gesture recognition, keystroke inference must distinguish among dozens of fine-grained key events that often exhibit subtle differences in their signal signatures. 
Third, generalization is a grand challenge. Typing patterns vary widely across users due to differences in finger strength, typing speed, and hand posture. Environmental factors such as desk material, ambient vibrations, and tag placement further compound this variability.

To address these challenges, we design a dedicated signal processing pipeline that extracts robust keystroke features from weak backscatter signals across both temporal and spectral domains, at both coarse and fine granularities.
In particular, the fine-grained feature design leverages the fact that two distinct signal paths exist from the keyboard to the RF tag:
(i) solid-borne typing vibrations, and (ii) air-borne acoustic sounds.
Leveraging their different propagation speeds, we compute the Time Difference of Arrival (TDoA) between these two signals as a discriminative, fine-grained feature for each keystroke. This feature is largely invariant to both user and keyboard, supporting strong generalization across different scenarios.
The coarse- and fine-grained features are fused via a cross-attention mechanism and passed to an offline-trained deep neural network (DNN) classifier to predict the typed content.

To further enhance \pname's adaptability in new attack scenarios, we introduce an LLM-guided online adaptation method. While prior studies \cite{ayati2025making, ni2024non, park2025improving, roh2025llm} use LLMs for post hoc text correction, \pname uses an LLM as a source of pseudo ground-truth labels to update the keystroke classifier. Prompting ensures that the online adaptation is triggered only for coherent, semantic text.
This mechanism effectively transfers the LLM's language knowledge into the keystroke classifier, significantly improving \pname's accuracy and adaptability even in challenging contexts such as non-semantic typing (e.g., passwords).

We have built a prototype of the RF tag and reader and evaluated \pname in various scenarios. 
Extensive experiments show that 
(i) \pname achieves robust and accurate keystroke inference across diverse keyboard types and user profiles,
(ii) the system remains effective under challenging conditions, such as through-wall and long-distance settings,
and 
(iii) the integration of LLM-guided online adaptation significantly improves generalization and reduces the character error rate without requiring victim-specific keystroke data for training.

This paper makes the following key contributions:
\begin{itemize}
\item 
We introduce a dual-resonator RF backscatter tag that can convert keystroke signals to the frequency shift of backscattered RF signals. More importantly, it separates the frequencies of excitation and backscatter signals and thus mitigates the self-interference issue. 

\item 
We design a novel RF reader that can extract reliable keystroke features from the backscatter signals. 
It features an LLM-guided online adaptation to improve its keystroke detection accuracy and adaptability. 

\item 
Extensive experiments confirm the effectiveness of this attack in diverse realistic scenarios. 

\end{itemize}

%% file: 3_threat_model_attack_overview.tex
\textbf{Attack Description.}
We consider the attack scenario as illustrated in Fig.~\ref{fig:sys_scenario}, where an adversary seeks to eavesdrop on keystrokes generated by a victim typing on a physical keyboard or a laptop. Based on the captured keystroke signals, the adversary aims to reconstruct the typed content and infer sensitive information such as messages, search queries, or passwords.
To perform this attack, the adversary discreetly places a passive RF tag beneath the surface of the desk or table supporting the keyboard. This setup leverages mechanical coupling, where keystroke-induced vibrations propagate through the rigid structure and are picked up by the tag, without requiring any direct contact with the keyboard itself.
The attack requires no modification to the keyboard or host device and remains effective as long as both the keyboard and tag are placed on the same rigid surface. 
This threat model is applicable to various environments, including offices, libraries, conference rooms, and co-working spaces.

\textbf{Assumptions.}
For this threat model, we make the following assumptions.
\textit{First}, we assume that a passive RF tag can be unobtrusively placed underneath the target desk or table. This is feasible in both public and private settings. In shared spaces such as offices, conference rooms, libraries, or cafes, one-time temporary physical access is often sufficient for discreet tag placement, allowing it to remain unnoticed for extended periods. In private spaces such as personal offices, this assumption also holds due to the risk of insider threats, such as cleaning staff, visitors, or disgruntled employees.
\textit{Second}, we assume the adversary knows the make and model of the victim's keyboard or laptop. This information can often be inferred through casual observation, public device usage patterns, or inventory information.
\textit{Third}, while the passive tag must be placed inside the target space, the RF reader may be located either within or outside the victim's physical environment, enabling both proximal and remote eavesdropping scenarios. This flexibility accounts for diverse real-world attack surfaces, including through-wall keystroke eavesdroppers or compromised on-site infrastructure.
We note that this attack is not undetectable. The passive backscatter tag may be discovered through visual inspection, and the RF eavesdropping activity can also be detected by spectrum monitoring.

%% file: 4_preliminary_study.tex
This section introduces the fundamental principles underlying our proposed keystroke eavesdropping attack.
Specifically, we examine how mechanical vibrations are generated by keystrokes, how these vibrations propagate through various media, and how they are detected by a piezoelectric sensor.

\subsection{Keyboard Switch Structure}
\begin{figure}
    \centering
    \includegraphics[width=1\linewidth]{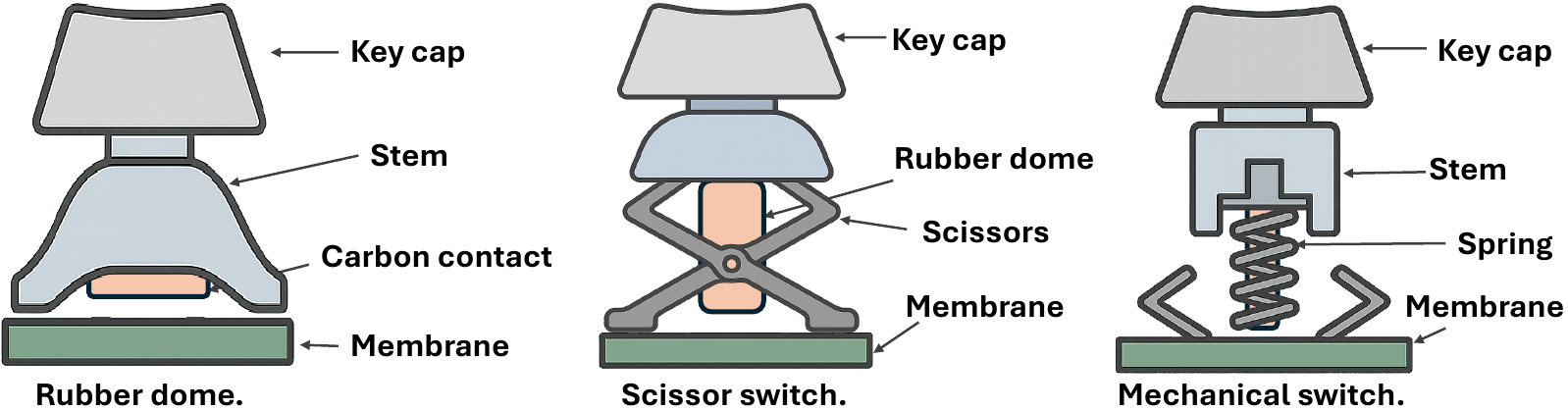}
    \caption{Three common keyboard switch types.}
    \label{fig:keys}
\end{figure}

Modern computer keyboards, despite their varied designs, primarily operate using three types of mechanical switch actuations: rubber dome switches, scissor-switch mechanisms, and mechanical switches \cite{monaco2018sok}. 
Each switch type exhibits distinct tactile feedback, actuation force, and acoustic profile, yet all produce impulse-like mechanical vibrations and acoustic emissions during keypresses due to the rapid contact and deformation events involved.
Fig.~\ref{fig:keys} illustrates the structural differences among these three common keyboard switch types.

\begin{itemize}[leftmargin=0.15in]
    \item \textit{Rubber dome switches}, commonly found in commodity keyboards, use a flexible rubber membrane that collapses under pressure to complete an electrical circuit. The deformation and subsequent rebound of the dome generate both acoustic sound and mechanical vibration.
    \item \textit{Scissor-switch keys}, typically used in laptops and low-profile keyboards, utilize a scissor-like structure to guide vertical key movement. Although designed for shorter travel and quieter operation, scissor switches still produce distinct vibrational and acoustic signatures due to their mechanical contacts.
    \item \textit{Mechanical switches} incorporate individual spring-loaded components beneath each key. Variants such as Cherry MX and Alps switches differ in actuation force and tactile response, but all share a common trait: physical component interactions (e.g., stem movement and contact closure) that produce both audible clicks and strong mechanical vibrations.
\end{itemize}

Regardless of switch types, pressing and releasing a key involves multiple distinct mechanical events, including finger contact with the keycap (touch peak), actuation at the bottom of the keystroke (hit peak), and return motion (release peak). 
Each of these events contributes to a complex waveform comprising both air-borne acoustic energy and mechanical vibrations that propagate through the keyboard and its supporting surface. 
These signals serve as the initial stimuli for our dual-path vibration sensing approach.

\begin{figure}
    \centering
    \includegraphics[width=1\linewidth]{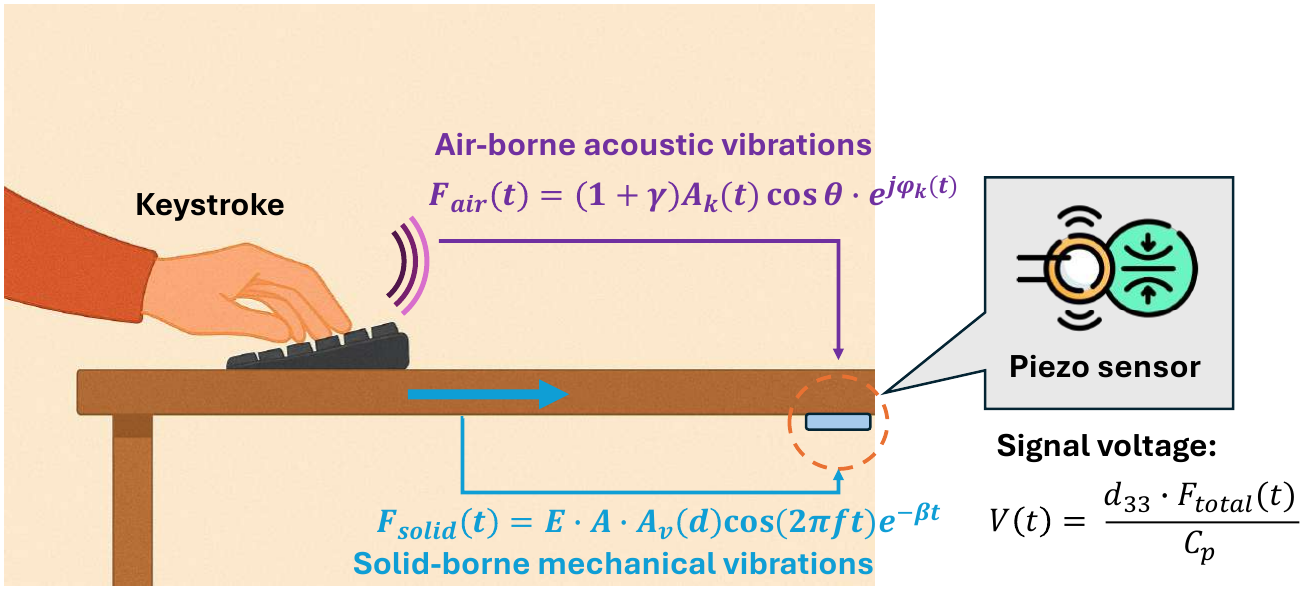}
    \caption{Keystroke-induced signal propagation model.}
    \label{fig:vib_model}
\end{figure}

\subsection{Keystroke-induced Object Vibration Model}
\label{object_vib}
When a key is pressed, the applied force initiates a mechanical impulse that excites vibrations in both the keyboard and its supporting surface, such as a table. These vibrations propagate outward from the impact point through two distinct transmission paths: solid-borne and air-borne propagation. 
Fig.~\ref{fig:vib_model} illustrates this dual-path propagation model.

\textbf{Solid-Borne Mechanical Vibration.}
In the solid-borne path, the mechanical energy of a keystroke generates damped longitudinal vibrations in the keyboard body and table surface. These vibrations decay in amplitude as they propagate through the solid material due to internal damping and scattering. The amplitude  $A_v(d)$ at a distance $d$ from the source can be expressed as \cite{graff2012wave}:
\begin{equation}
    A_v(d) = A_{v0} e^{-\alpha d},
\end{equation}
where $A_{v0}$ is the initial amplitude, and $\alpha$ is the attenuation coefficient determined by the physical properties of the medium and vibration frequency.

To estimate the force experienced by the sensor, we assume the local displacement follows a damped sinusoidal form \cite{thomson2018theory}:
\begin{equation}
u(t) = A_v(d) \sin(2\pi f t) e^{-\beta t},
\end{equation}
where $f$ represents the dominant resonant frequency excited by the keystroke, and $\beta$ is the temporal damping coefficient. This waveform reflects the transient nature of keypresses.

The strain induced by this vibration is the spatial derivative $\partial u(t)/\partial d$, which produces a time-varying stress $\sigma(t)$ according to Hooke’s law for linear elastic solids \cite{landau2012theory}:
\begin{equation}
\sigma(t) = E \frac{\partial u(t)}{\partial d},
\end{equation}
where $E$ is the Young’s modulus of the surface material. The total dynamic force applied to the piezoelectric sensor area $A$ is \cite{timoshenko2012theory}:
\begin{equation}
    F_{solid}(t) = \sigma(t) A = -\alpha \cdot E \cdot A \cdot A_v(d) \cdot \sin(2\pi f t) e^{-\beta t}.
\end{equation}

This force serves as the sensor input, leading to voltage generation as described in Section~\ref{piezo}.

\textbf{Air-Borne Acoustic Sound.}
Concurrently, the keystroke produces air-borne acoustic waves that travel through the surrounding air. These waves, characterized by amplitude $A_k(t)$ and phase $\phi_k(t)$, induce surface pressure. The total acoustic pressure $F_{air}(t)$ at the surface, incorporating both direct and reflected components, is given by \cite{kinsler2000fundamentals}:
\begin{equation}
    F_{air}(t) = (1+\Gamma) A_k(t) \cos\theta \cdot e^{j\phi_k(t)},
\end{equation}
where $\Gamma$ is the reflection coefficient based on surface impedance, and $\theta$ is the angle of incidence of the wave.

\begin{figure}
    \centering
    \includegraphics[width=0.99\linewidth]{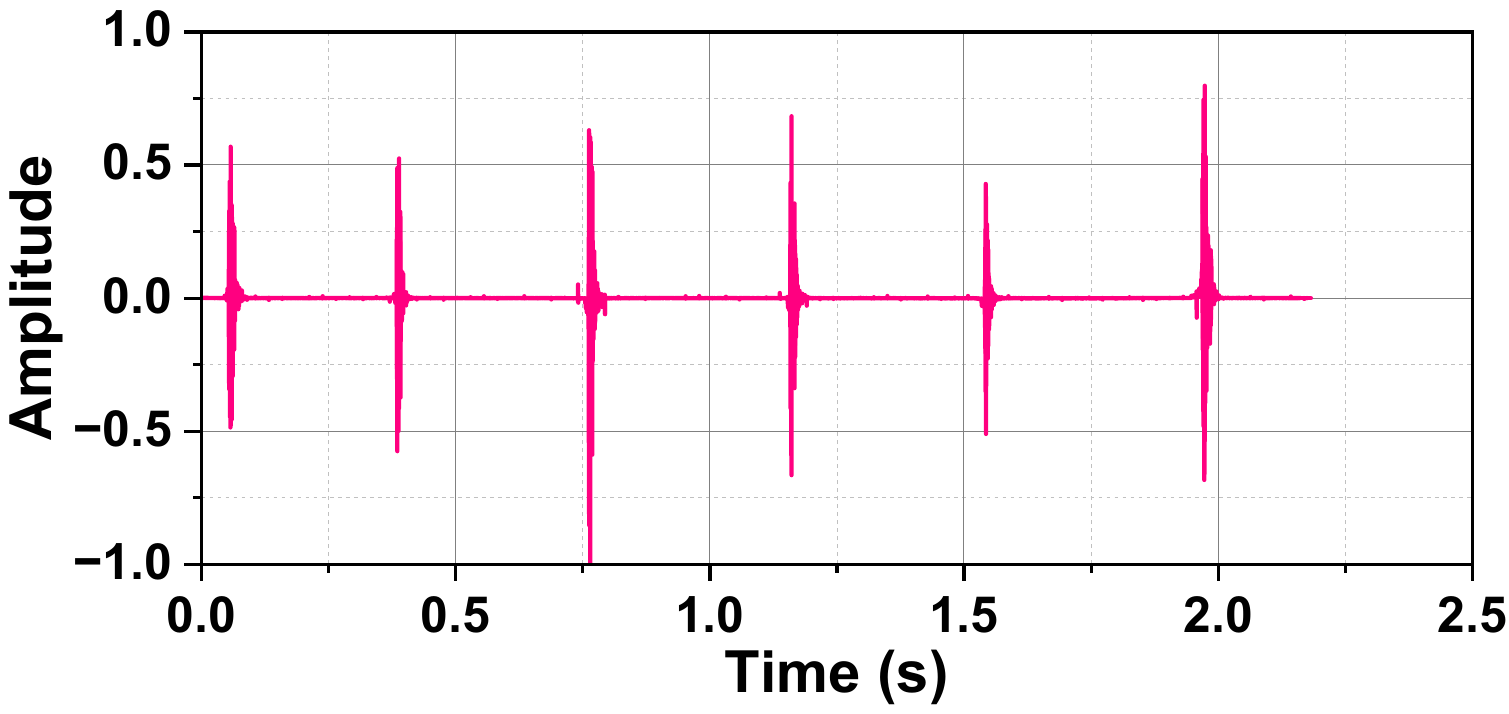}
    \caption{Demodulated keystroke signal captured when the keyboard and RF tag are placed on separate tables.}
\label{fig:dif_tables}
\end{figure}

\subsection{Piezoelectric Sensor for Keystroke Detection}
\label{piezo}
Piezoelectric sensors are widely used for detecting mechanical vibrations due to their ability to convert dynamic stress into electrical voltage. 
This property arises from the \emph{piezoelectric effect}, wherein certain crystalline materials generate an electric charge when subjected to mechanical deformation.

When a time-varying force $F(t)$ acts on a piezoelectric material, it induces an electrical potential difference across the sensor terminals. 
In our keystroke-induced vibration model, the total force acting on the sensor is the superposition of forces from solid-borne and air-borne vibrations, i.e., 
\begin{equation}
    F(t) = F_{\text{solid}}(t) + F_{\text{air}}(t). 
\end{equation}
This combined mechanical input drives the sensor response and determines the resulting voltage output.
The corresponding output voltage $V(t)$ is modeled as \cite{uchino1996piezoelectric}:
\begin{equation}
    V(t) = \frac{d_{33} }{C_p} \cdot F(t),
\end{equation}
where $d_{33}$ is the charge coefficient characterizing the material's response,  $C_p$ is the internal capacitance of the sensor, and $F(t)$ is the mechanical force applied due to surface vibrations.

Due to their high sensitivity and broad frequency response, piezoelectric sensors can respond to both solid-borne and air-borne vibrations. 
To validate this capability, we perform an isolation experiment where the keyboard and the RF tag are placed on separate tables with no mechanical coupling. 
Despite the separation, the tag reliably captures keystroke-induced air-borne signals, as shown in Fig.~\ref{fig:dif_tables}.
This confirms that the piezoelectric sensor is sufficiently sensitive to air-borne acoustic energy in realistic typing scenarios. 
It therefore enables robust dual-path signal acquisition in our system.

Since the mechanical input force is directly related to the vibration dynamics outlined in Section~\ref{object_vib}, the resulting voltage signal $V(t)$ contains temporal and spectral information about the original keystroke event. For small-amplitude, linear deformations, the sensor's response remains proportional to the local stress field or surface acceleration.


\begin{figure}
    \centering
    \includegraphics[width=0.99\linewidth]{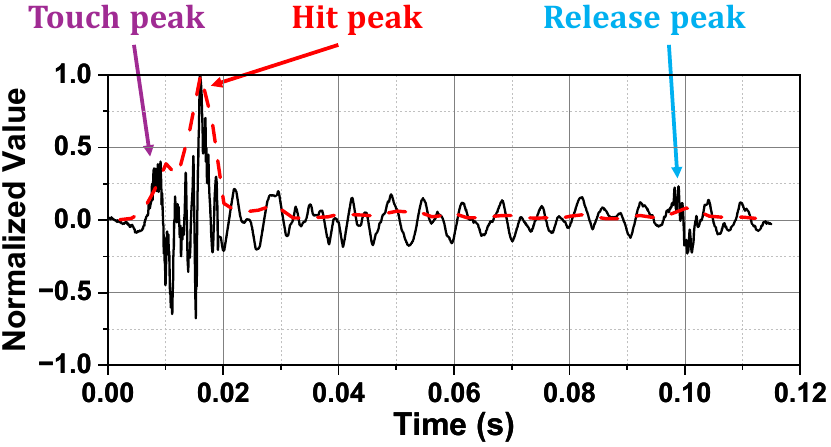}
    \caption{Example of keystroke signal captured by a piezoelectric sensor.}

\label{fig:peaks}
\end{figure}

Fig.~\ref{fig:peaks} shows a signal sample captured by a piezoelectric sensor during a single keystroke event. The waveform reveals three prominent peaks corresponding to distinct mechanical interactions: the initial finger contact with the keycap, the bottom-out impact at full depression, and the elastic rebound as the key returns to its rest position. These peaks differ in amplitude and spacing depending on typing style and key structure, but consistently follow a characteristic temporal order. The sensor output reflects the superposition of solid-borne and air-borne components induced by the keystroke, resulting in a rich, non-stationary waveform containing both high- and low-frequency content critical for downstream inference.

%% file: 5_RFtag_based_vibration_acquisition_and_feature_extraction.tex
\subsection{Tag Design}

The RF backscatter tag plays a critical role in enabling reliable keystroke eavesdropping.
First, the tag must be compact (e.g., less than one inch in diameter) and easily concealable within typical target environments.
Second, it must support continuous operation powered solely by energy harvesting.
Most prior RF backscatter systems, such as WISP~\cite{menon2022wireless} and MARS~\cite{arora2021mars}, cannot sustain continuous communication and instead require prolonged energy harvesting periods, resulting in low-duty-cycle and intermittent operation.
Third, it is highly desirable for the tag to decouple its excitation and reflection frequencies, so as to avoid self-interference at the RF reader.

\begin{figure}
    \centering
    \includegraphics[width=0.8\linewidth]{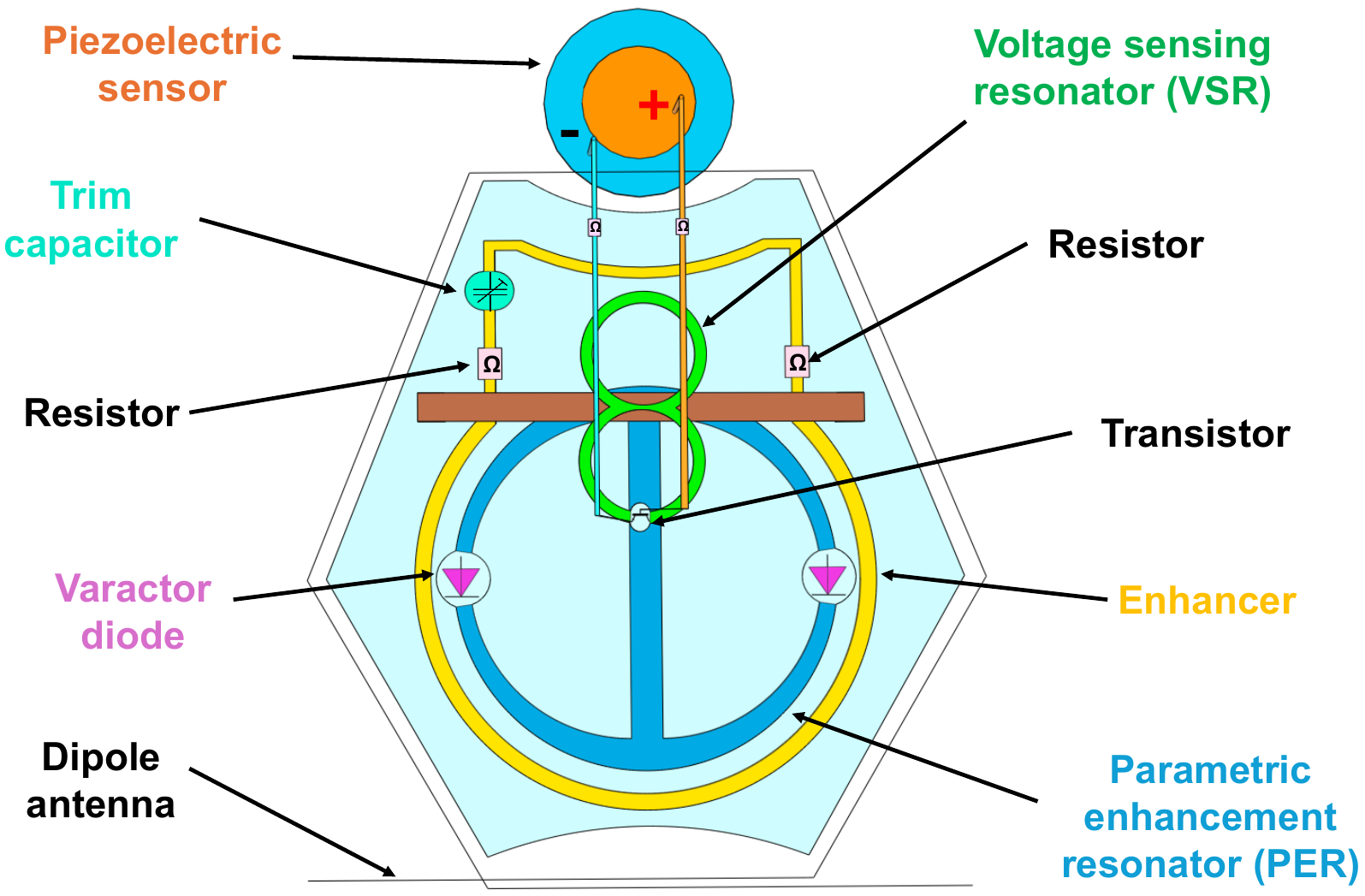}
    \caption{The diagram of our proposed backscatter tag.
    }
    \label{fig:tag_design}
\end{figure}

\textbf{Tag Structure.}
To meet these requirements, we propose a dual-resonator tag as shown in Fig.~\ref{fig:tag_design}, which is composed of four key components:
(i) a piezoelectric sensor,
(ii) a voltage sensing resonator (VSR),
(iii) a parametric enhancement resonator (PER), and
(iv) a dipole antenna.
An off-the-shelf piezoelectric sensor that is sensitive to both solid-borne mechanical vibrations and air-borne acoustic sound is selected for this tag.
The VSR is an LC circuit composed of an 8-shaped coil terminated by a bipolar junction transistor (BJT) whose base and emitter connect to the piezoelectric sensor.  
The PER is a circular planar resonator on a fiberglass-epoxy substrate incorporating two varactor diodes across split gaps to support dual-mode resonance (circular and butterfly modes).

 \textbf{Operation Principles.}
The piezo sensor captures the keystroke-induced vibrations and acoustic fluctuations and converts them into a voltage signal, which modulates the capacitance of a diode in the VSR.
Since the resonance frequency of VSR depends on this capacitance, the VSR directly translates the voltage signal into frequency shifts, enabling frequency modulation of the captured sound and vibration.
The PER is magnetically coupled to the VSR.
Together with the dipole antenna, the PER serves two purposes:
(i) harvest energy from the RF reader to sustain the resonance, and
(ii) enhance radiation efficiency to strengthen the backscattered signal.
More importantly, the PER separates its excitation and reflection frequencies by adopting a dual-mode resonance structure. 
This design mitigates self-interference at the RF reader, thereby extending the effective eavesdropping range.

\subsection{Signal Analysis}
\textbf{VSR Operation.}
For notational simplicity, denote $v_s$ as the piezoelectric sensor's instantaneous voltage output, which is proportional to the mechanical force induced by keystrokes. 
This voltage is applied to the varactor diodes of the VSR, whose capacitance varies with $ v_s$. 
Denote $f_1$ as the resulting resonance frequency of the VSR.
Then, we have:
\begin{equation}
f_1 = \frac{1}{2\pi}\sqrt{\frac{2}{L_1 C_1}} = \frac{1}{2\pi} \sqrt{\frac{2}{L_1 C_{10} \left(1 - \frac{v_s}{\Phi_1} \right)^{-\lambda_1}}},
\label{eq:vsr_1}
\end{equation}
where $L_1$ and $C_1$ are the VSR's inductance and capacitance, respectively; $C_{10}$ is its capacitance when the bias voltage is 0.
$ \Phi_1 $ is the varactor diode's junction potential. 
$\lambda_1$ is a constant related to the property of the varactor diode. 

In practice, we have $ v_s \ll \Phi_1 $. 
Based on the first-order Taylor approximation, we have: 
\begin{equation}
f_1 \approx  f_{10} \left(1 - \frac{\lambda_1 v_s}{\Phi_1} \right), 
\label{eq:vsr_2}
\end{equation}
where $f_{10} \equiv \frac{1}{2\pi} \sqrt{\frac{2}{L_1 C_{10}}}$ is the unperturbed resonance frequency of the VSR when no bias voltage is applied.
This establishes a linear mapping between the keystroke-induced voltage and the carrier frequency shift, enabling frequency modulation in the analog domain without digitization.

\textbf{PER Operation.}
The PER is introduced to enhance the backscatter signal strength so as to extend the eavesdropping range. 
The PER is designed to operate in two resonance modes simultaneously: \textit{circular mode} and \textit{butterfly mode}. 
This is achieved via a symmetric half-circle layout with dual varactor diodes as shown in Fig.~\ref{fig:tag_design}. 
The circular mode couples magnetically to the VSR and thus captures frequency shifts caused by keystrokes, while the butterfly mode supports excitation by the RF reader.

\begin{figure}
    \centering
    \includegraphics[width=0.6\linewidth]{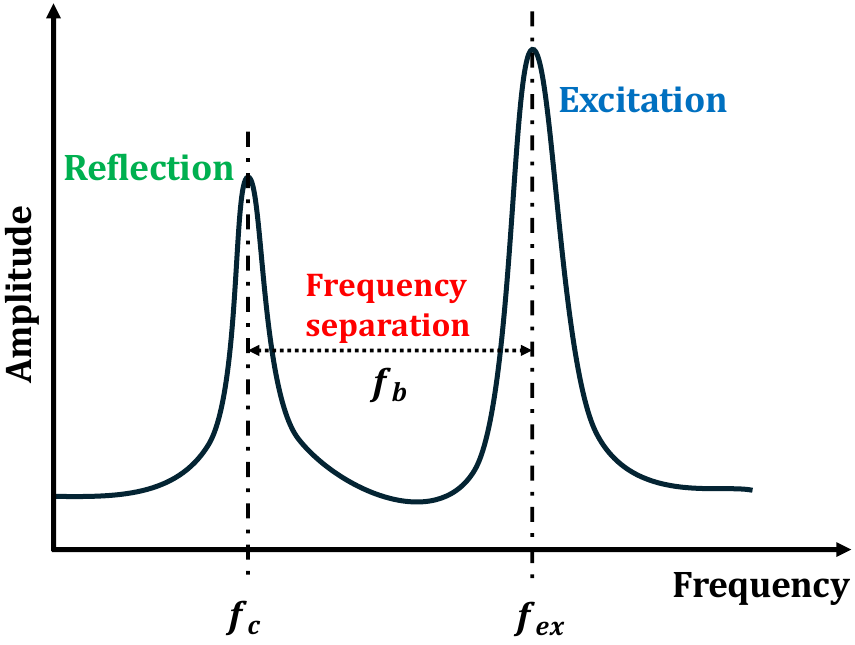}
    \caption{
Illustrating spectral separation of the RF tag's excitation and reflection/backscatter signals.
}
    \label{fig:freq_sep}
\end{figure}

Denote $f_c$ as the resonance frequency of its circular mode. 
Denote $f_b$ as the resonance frequency of its butterfly mode. 
When the PER is excited by a signal at frequency $f_{\text{ex}} = f_c + f_b$, it will radiate the modulated backscatter signal at frequency $f_c$, which we call either the reflection frequency or the backscatter frequency. 
Fig.~\ref{fig:freq_sep} illustrates the separation of reflection and excitation frequencies. 
This fundamentally mitigates the self-interference issue for the RF reader and thus increases the eavesdropping range. 


\textbf{Tag Modulation Analysis.}
Recall that $v_c$ is the voltage generated by the piezo sensor and $f_c$ is the tag's backscatter signal frequency.
Then, we have the following theorem:
\begin{theorem}
\label{therem:fm}
\em
The proposed tag achieves a frequency modulation for the voltage signal generated by the piezo sensor, i.e., 
\begin{align}
\frac{\partial f_c}{\partial v_s} = 
& \bigg( -\frac{\lambda_1 f_{10}}{2 \Phi_1} \bigg) 
\bigg( \frac{\frac{L_c}{R_c}}{\frac{L_c}{R_c} + \frac{L_b}{R_b}} \bigg)
\bigg( \frac{f_L^3}{2 f_1^3} \bigg)  
\nonumber\\
&
\underbrace{\bigg( 1 + \frac{\frac{f_2^2}{f_1^2} - 1 + 2 \left( \frac{f_1^2}{f_L^2} - 1 \right) \left( \frac{f_2^2}{f_L^2} - 1 \right)}{\left| 2 \frac{f_2^2}{f_L^2} - \frac{f_2^2}{f_1^2} - 1 \right|} \bigg)}_{\textbf{Constant}}
,
\label{eq:linear_freq_vs}
\end{align}
where 
$f_1$ and $f_2$ are the stand-alone resonance frequencies of VSR and PER, respectively. 
$f_L$ is the lower resonance frequency of the coupled VSR and PER resonators. 
$R_c$ and $L_c$ are the effective resistance and inductance of the PER in its circular mode.
$R_b$ and $L_b$ are the effective resistance and inductance of the PER in its butterfly mode.
\end{theorem}

The proof is provided in Appendix~\ref{app:theorem:fm}.
Theorem~\ref{therem:fm} shows that the frequency of the tag's backscattered radio signal varies linearly with the output voltage of the piezoelectric sensor, confirming the tag's \textit{frequency modulation} (FM) behavior.
We note that FM is widely regarded as a more reliable modulation scheme than alternatives such as amplitude modulation (AM).


%% file: 6_LLM-powered_privacy_information_inference.tex
To demodulate the keystroke signal and estimate the victim's typing content, we propose an RF reader design composed of three key components:
(i) backscatter signal demodulation, 
(ii) keystroke feature extraction, 
and
(iii) LLM-guided online adaptation.
We explain them below.

\subsection{Backscatter Signal Demodulation}


\begin{figure}
\centering
\includegraphics[width=1\linewidth]{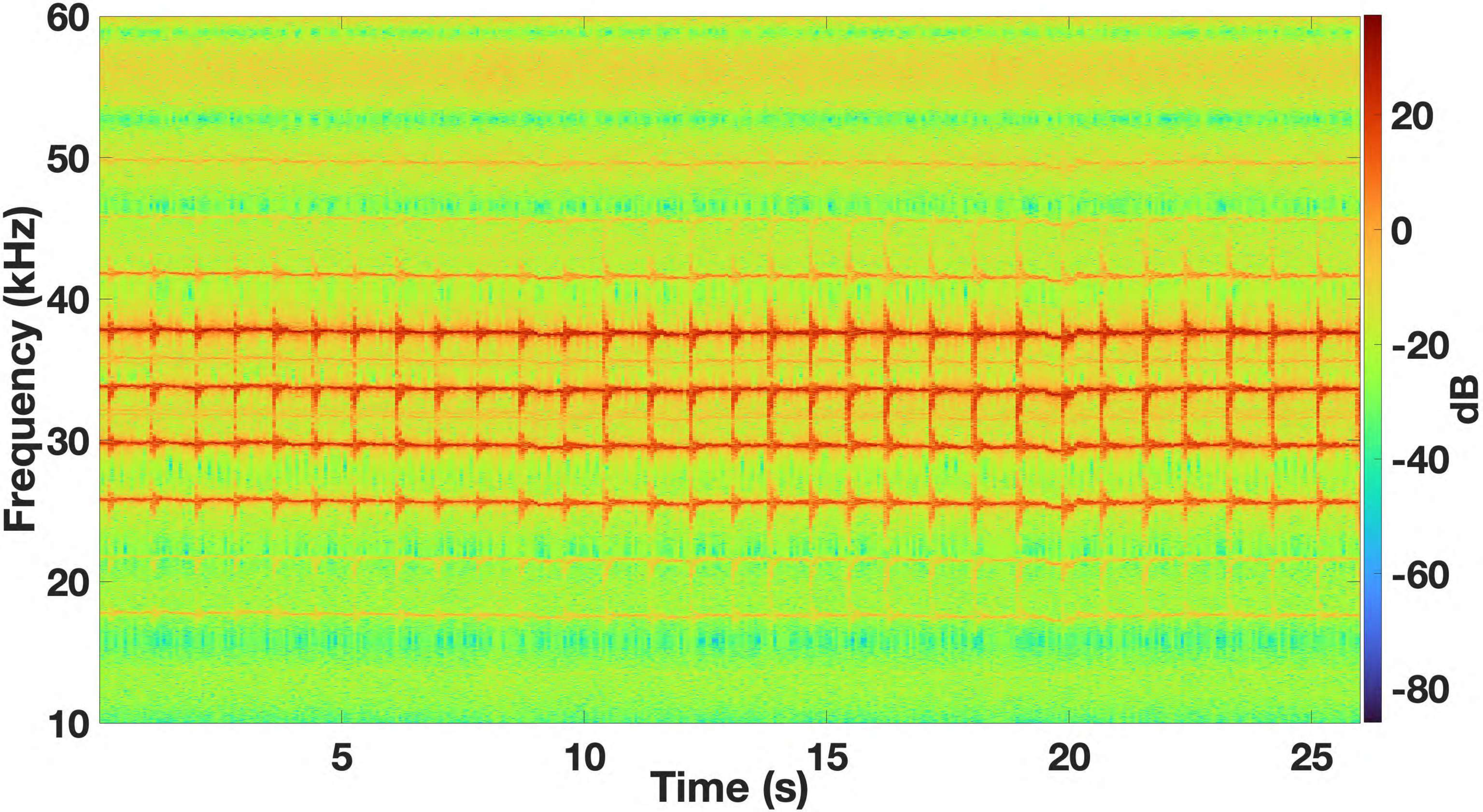}
\caption{
Spectrogram of the received multi-carrier backscatter signal at the RF reader. 
}
\label{fig:multi_freq}
\end{figure}

\textbf{Backscatter RF Signal.}
Due to imperfections in tag fabrication, the keystroke voltage signal is modulated onto multiple backscatter carriers rather than a single backscatter carrier.
Specifically, the received radio signal at the RF reader can be modeled as:
\begin{equation}
r(t) = \alpha \sum_{k=1}^{K} \cdot \exp\big(2\pi f_{k} t + 2\pi \Delta_k \int_0^t v_s(\tau) d\tau\big) + w(t),
\label{eq:multi_fm}
\end{equation}
where \( f_{k} \) represents the center frequency of the \( k \)-th carrier, $K$ is the total number of carriers, 
$\Delta_k$ is the frequency modulation coefficient of carrier $k$, $\alpha$ is the signal attenuation coefficient shared by all carriers, and \( w(t) \) denotes additive noise and imperfection artifacts.
We note that this model does not account for the multi-path effect of RF signal propagation.
This is because the frequency-modulated keystroke signal on the RF backscatter carrier is very narrowband (less than $40\,\text{kHz}$) and thus the multi-path effect is insignificant, especially for frequency modulation.

Fig.~\ref{fig:multi_freq} shows an example of the received signal at the RF reader. 
Clearly, the received signal contains multiple carriers with different strengths. Each carrier is frequency-modulated by the keystroke voltage signal.
The carriers are equally spaced. 
To demodulate the keystroke voltage signal, we simply keep the strongest carrier and remove all other carriers. 
This method is not optimal, but our experiments show that its performance loss is marginal.



\textbf{CFO Correction.}
The frequency-modulated backscatter signal is first down-converted to an intermediate frequency (IF) (e.g., 40~kHz) to avoid DC interference and preserve low-frequency keystroke content. 
However, due to tag hardware imperfections and environmental factors (e.g., temperature drift), the center frequencies of reflection signals are not static but drift slowly over time. 
These variations, referred to as carrier frequency offsets (CFO), degrade FM demodulation performance if uncorrected.

To address this, we implement a CFO estimation and compensation module in the frequency domain. 
The incoming signal is segmented into $N$-point frames, and each segment undergoes Fast Fourier Transform (FFT) to locate the center frequency of the strongest carrier. 
Let $f_{c_0}$ be the estimated center frequency of the dominant carrier used for demodulation. 
Then, the offset $\Delta f$ is calculated for each segment and applied to shift the IF signal to baseband (zero-IF), i.e.,
$\hat{x}(t) = x(t) \cdot e^{-j2\pi \Delta ft}$, where $x(t)$ is the sampled IF signal, and $\hat{x}(t)$ is the CFO-compensated baseband signal.

\textbf{Signal Demodulation.}
After the CFO has been corrected, the RF reader is ready to demodulate the keystroke signal using phase differentiation. 
Let $\hat{x}(t)$ denote the complex-valued baseband signal after low-pass filtering. Then, the instantaneous phase is given by:
$ \phi(t) = \angle \hat{x}(t) = 2\pi \Delta \int_0^t v_s(\tau) d\tau $.
The keystroke signal $v_s(t)$ can be estimated by differentiating $\phi(t)$ over time:
\begin{equation} 
v_s(t) \approx \frac{f_s}{2\pi \Delta} \left[\phi(t+\Delta t) - \phi(t)\right],
\label{eq:vs_out}
\end{equation}
where $f_s$ is the sampling rate, $\Delta$ is the frequency modulation coefficient of the selected backscatter carrier used for demodulation, and $\Delta t$ is a small step size (one sample interval). 
After demodulation, the demodulated keystroke signal is resampled to a uniform rate (e.g., 44.1~kHz) for feature extraction.



\begin{figure}
    \centering
    \includegraphics[width=\linewidth]{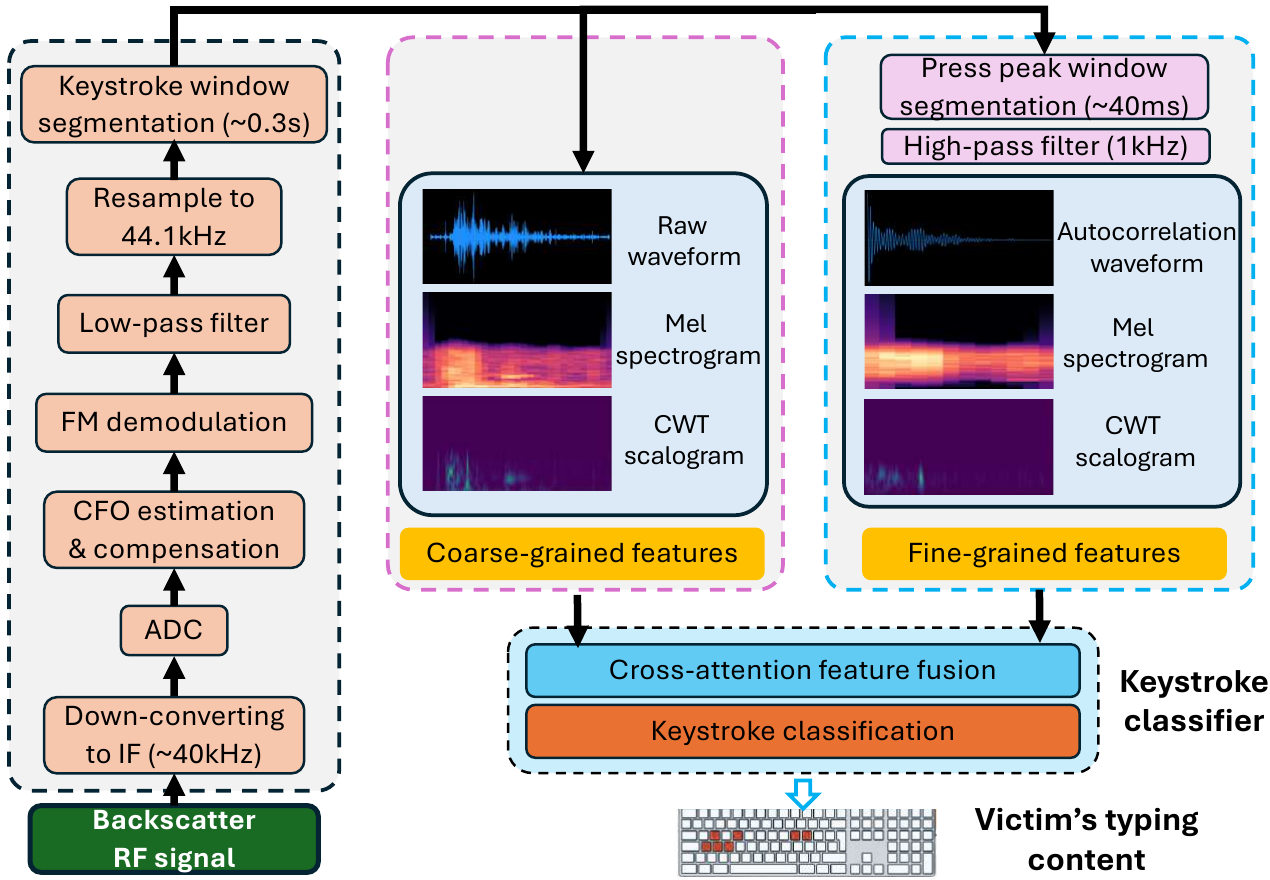}
    \caption{Overview of RF reader's signal processing and feature extraction.}
    \label{fig:feature_extraction}
\end{figure}

\subsection{Keystroke Signal Segmentation}
To enable reliable inference, \pname requires a lightweight yet robust keystroke segmentation mechanism that can isolate individual keystroke events from continuous backscattered signals. We adopt a self-adaptive, energy-based segmentation strategy that avoids dependency on fixed thresholds or complex model-based post-alignment.

\textbf{Short-Time Energy Envelope.}
Denote $s[n]$ as the digitalized samples of the demodulated keystroke signal $v_s(t)$ in Eqn~\eqref{eq:vs_out}.
Then, we compute its short-time energy envelope $A[n]$ using a sliding window of length $L$:
\begin{equation}
A[n] = \sqrt{ \frac{1}{L} \sum_{k=0}^{L-1} s^2[n - k] }.
\end{equation}
This operation captures transient energy bursts caused by keystroke events while suppressing background variations.

\textbf{Self-Adaptive Thresholding.}
Instead of using a fixed threshold, we compute a dynamic threshold using local signal statistics.
A local threshold is computed at each time index based on the surrounding energy distribution. 
Specifically, for each point $n$, we compute a local mean $\mu[n]$ and standard deviation $\rho[n]$ of $A[n]$ over a sliding window, and mark $n$ as a keystroke onset if:
$A[n] > \mu[n] + \lambda \cdot \rho[n]$,
where $\lambda$ is a tunable sensitivity factor. This formulation enables the segmentation to adapt to different signal amplitudes, background levels, and device placements.

\textbf{Onset Filtering and Segment Extraction.}
To eliminate spurious detections caused by reverberation or hand tremors, we enforce a minimum temporal spacing $\Delta_{\mathrm{min}}$ between adjacent onsets. Within each window of length $\Delta_{\mathrm{min}}$, only the peak with the highest energy is retained. Around each remaining onset time $n_i$, we extract a fixed-length segment $s_i[n]$ from the original signal, centered at $n_i$ with asymmetric margins.

\subsection{Two-Time-Scale Feature Extraction}

To effectively capture the rich information embedded in keystroke acoustic signals, we propose a hierarchical time-frequency feature extraction framework. This approach extracts and processes features at two complementary levels:
(i) a \textit{coarse-grained} level that characterizes the entire keystroke event, and 
(ii) a \textit{fine-grained} level that focuses on the detailed properties of keystroke peaks and their inter-channel timing relationships. Features from both levels are then fused and jointly processed to produce the final keystroke classification.

\textbf{Coarse-Grained Feature Extraction.}
We define a coarse-grained feature extraction stage that operates over the entire keystroke event, using a time window of approximately 330 ms. This duration is not fundamental to the method; rather, it is an empirically chosen upper bound that is sufficient to capture the full acoustic signature of a keystroke, including the initial touch, the primary impact (hit peak), and the subsequent release and reverberations.

From this time window, we extract the following three representative features.

\begin{itemize}[leftmargin=0.15in]
\item 
\textit{Raw Time-Domain Signal:} It preserves the temporal waveform morphology, which may encode unique patterns per key and user. 

\item 
\textit{Mel Spectrogram:} It converts the signal into a time-frequency representation emphasizing perceptual frequency resolution, mimicking human auditory processing.

\item 
\textit{Wavelet Transform:} It provides excellent time-frequency localization, adept at capturing transient events and varying frequency components for the keystroke.
\end{itemize}

These three representations are then fed into a dedicated coarse-grained feature extraction sub-network implemented using a lightweight CoAtNet \cite{dai2021coatnet} architecture. 
CoAtNet combines the strengths of convolutional operations for local feature extraction and attention mechanisms for modeling long-range dependencies, offering both efficiency and representational power. 
This sub-network learns abstract, high-level features that characterize the global structure of each keystroke event across input modalities. 
The outputs from processing each representation are subsequently concatenated for downstream fusion and classification.


\textbf{Fine-Grained Feature Extraction.}
The fine-grained level focuses on extracting precise details from the critical peak events within the keystroke (solid-borne and air-borne peaks) and robustly encoding the Time Difference of Arrival (TDoA) between the two channels.
Instead of using the entire 330~ms window, we define a shorter analysis window centered around the detected solid-borne hit peak. 
This window is extended to reliably encompass the later-arriving air-borne hit peak, considering the physical dimensions of the setup. 
Let $D_{table}$ be the longest dimension of the table (or relevant surface) supporting the keyboard.
Then, the time difference for sound propagation in air across this dimension is $D_{table}/c_{sound}$, where $c_{sound}$ is the sound speed (approx. 340~m/s). 
The fine-grained window is therefore defined to capture the solid-borne hit peak and this subsequent air-borne acoustic information, plus a small margin for variability (e.g., $duration\_solid\_peak + D_{table}/c_{sound} + padding$).

A crucial fine-grained feature is TDoA. 
Direct TDoA calculation using cross-correlation on raw signals can be highly sensitive to noise, multipath interference, and environmental factors. 
To create a more robust representation, we propose transforming the TDoA information into an image-like feature. 
This is achieved by applying time-frequency transformations (e.g., Mel spectrograms or CWTs) to the segmented audio within this fine-grained window.
The resulting image-based time-frequency representation inherently captures the phase and arrival time differences.
It serves as an ``image" where the TDoA is encoded in the subtle misalignments and structural differences between them. 
This distributed representation is expected to be more resilient to noise and other imperfections than a single scalar TDoA value.
These time-frequency representations (TDoA images) are then processed by a separate fine-grained feature extraction sub-network.
This network learns to extract discriminative features from these ``TDoA images" that are indicative of the precise inter-channel timing differences and specific peak characteristics.

\begin{figure}
    \centering
    \includegraphics[width=1\linewidth]{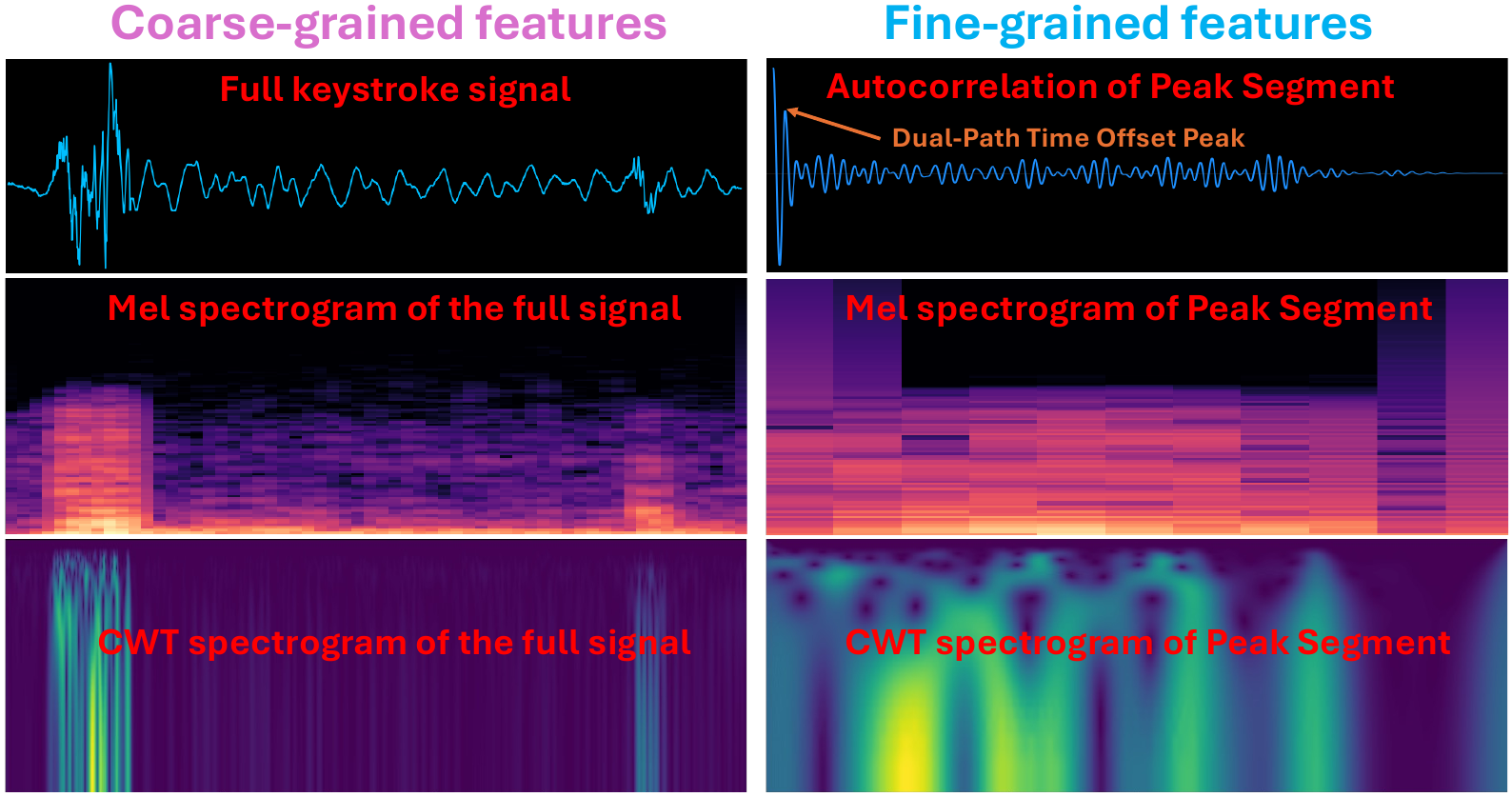}
    \caption{
    A sample of coarse-grained and fine-grained features.
    }
    \label{fig:features}
\end{figure}

Fig.~\ref{fig:features} shows an example of the coarse-grained and fine-grained features used in our two-scale feature extraction pipeline. These features serve as the foundational inputs for downstream keystroke classification, capturing both global event characteristics and transient peak-level details.

\textbf{Feature Fusion and Keystroke Classification.}
As illustrated in Fig.~\ref{fig:feature_extraction}, the features extracted from the coarse-grained and fine-grained pathways provide complementary information about each keystroke. The high-level vectors from the coarse-grained analysis (capturing the overall event) and the fine-grained analysis (capturing peak details and TDoA) are concatenated to form a unified feature representation. This fused vector encodes both the long-term spectral–temporal structure of the full keystroke and the transient characteristics around the peak regions.
The concatenated features are then processed by a cross-attention mechanism, which allows the model to dynamically assess the relative importance of the global event characteristics and the detailed temporal features. By learning to selectively emphasize the most discriminative elements within the combined feature space, the cross-attention layer produces a refined and contextually weighted representation. Finally, this output is passed to a classification head to predict the keystroke class.

\textbf{Offline Training.}
We collect keystroke data from multiple participants who are instructed to type predefined content consisting of both isolated letters and complete sentences. This mixture allows us to capture the natural variability in keystroke dynamics, as single-letter inputs and sentence-level typing often exhibit different temporal and spectral characteristics due to variations in inter-key intervals, finger transitions, and contextual hand movements.
The recorded signals are processed using the pipeline as shown in Fig.~\ref{fig:feature_extraction}, which includes acoustic signal recovery, keystroke segmentation, and feature extraction. The known ground-truth labels from the predefined input are then used to supervise the training of the keystroke inference model.

\subsection{LLM-Guided Online Adaptation}

\begin{figure}
    \centering
    \includegraphics[width=0.85\linewidth]{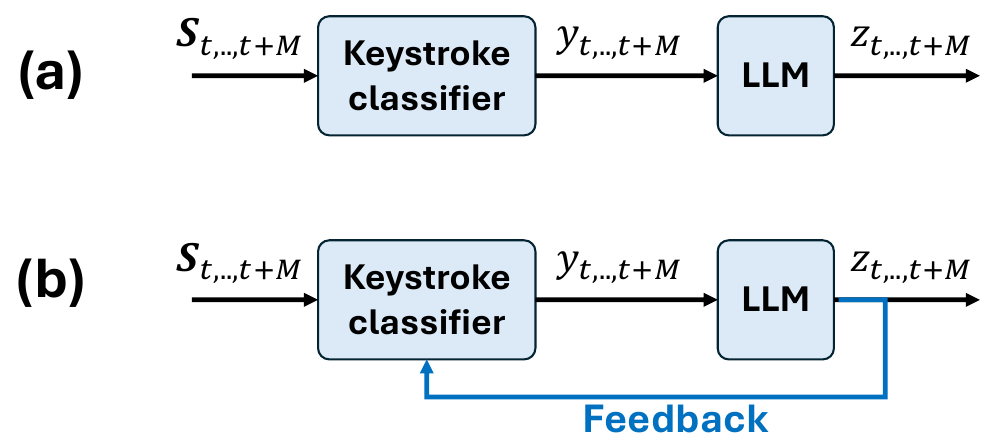}
    \caption{(a) LLM is used only for post hoc correction; and (b) LLM is used for online adaptation.}
    \label{fig:llm}
\end{figure}

\textbf{Main Ideas.}
LLMs are powerful tools for contextual text correction, and recent work \cite{ayati2025making, ni2024non, park2025improving, roh2025llm, lin2026integrating, lu2026eexapp, yan2025near, yan2025xdiff} has explored integrating them into keystroke inference pipelines. However, existing approaches use LLMs only as post-processors, as illustrated in Fig.~\ref{fig:llm}(a).
Specifically, the raw or top-k outputs from a keystroke classifier are fed into an LLM, which is prompted to refine them using its internal language knowledge. While this can improve readability and lexical plausibility, it restricts the LLM's role to surface-level correction and leaves the underlying classifier unchanged. As a result, such approaches may enhance eavesdropping performance for coherent text but offer little benefit for non-semantic inputs such as passwords.

To overcome this limitation, we propose an LLM-guided online adaptation framework for \pname, as shown in Fig.~\ref{fig:llm}(b). In our design, the LLM is not a standalone post-processor. Instead, when the victim types coherent text, the LLM corrects the classifier's output and treats the corrected text as pseudo ground-truth labels to supervise online updates of the keystroke classifier. In this way, language knowledge from the LLM is transferred into the classifier itself, enabling accuracy improvements even when the LLM is no longer involved.

Crucially, the keystroke classifier and the LLM operate at different levels: the classifier predicts individual keystrokes or characters, whereas the LLM models relationships across entire sentences. Through online adaptation, the classifier’s predictions improve even for non-semantic inputs such as passwords.
To ensure safe and effective adaptation, the LLM must trigger online training only when the input is coherent. A natural question is how the system determines coherence. \pname leverages the LLM's inherent language modeling ability: by prompting the LLM to evaluate the semantic plausibility of the classifier’s output, the system initiates online adaptation only when the input is contextually meaningful. This safeguard prevents misadaptation on random, password-like, or otherwise incoherent inputs, ensuring that updates remain grounded in valid language patterns.

\textbf{Our Design.}
Fig.~\ref{fig:llm}(b) shows our online adaptation framework.
Let $f_\theta(\cdot)$ be the keystroke classifier parameterized by $\theta$.
Let $\mathcal{S}_t$ be the set of coarse-grained and fine-grained features corresponding to one keystroke. 
Then, we define a loss function as:
\begin{equation}
\mathcal{L}_{\text{LLM-Align}} = \sum_{t} \mathcal{L}_{\text{CE}}\Big(f_\theta(\mathcal{S}_t), \; \operatorname{LLM}(f_\theta(\mathcal{S}_t)) \Big),
\end{equation}
where $\mathcal{L}_{\text{CE}}(\cdot, \cdot)$ is the cross-entropy loss function, $\operatorname{LLM}(\cdot)$ is the output of the LLM. 
This loss function uses the LLM to supervise the training of the keystroke classifier, leveraging the LLM's semantic refinement to improve the overall keystroke estimation accuracy.



Since the output of LLM is not always correct, we use a hybrid loss function, instead of solely cross-entropy loss $\mathcal{L}_{\text{LLM-Align}}$, as the loss function to update the keystroke classifier.
Specifically, the hybrid loss function is defined as:
\begin{equation}
\mathcal{L}_{\text{total}} = \lambda_1 \mathcal{L}_{\text{LLM-Align}} 
+ \lambda_2 \mathcal{L}_{\text{Conf}}(f_\theta(\mathcal{S}_t)) 
+ \lambda_3 \mathcal{L}_{\text{Smooth}}(\theta),
\end{equation}
where $\mathcal{L}_{\text{Conf}}$ is the confidence penalty to discourage overconfident incorrect predictions (entropy regularization),
$\mathcal{L}_{\text{Smooth}}$ is parameter smoothing to avoid catastrophic forgetting,
$\lambda_1$, $\lambda_2$, $\lambda_3$ are empirical weights balancing supervision, regularization, and stability.
This hybrid loss function promotes both the alignment with LLM-inferred labels and stabilizes the convergence under noisy updates.




%% file: 7_evaluation.tex

\subsection{Implementation}
\begin{figure}[t]
    \centering
    \includegraphics[width=0.99\linewidth]{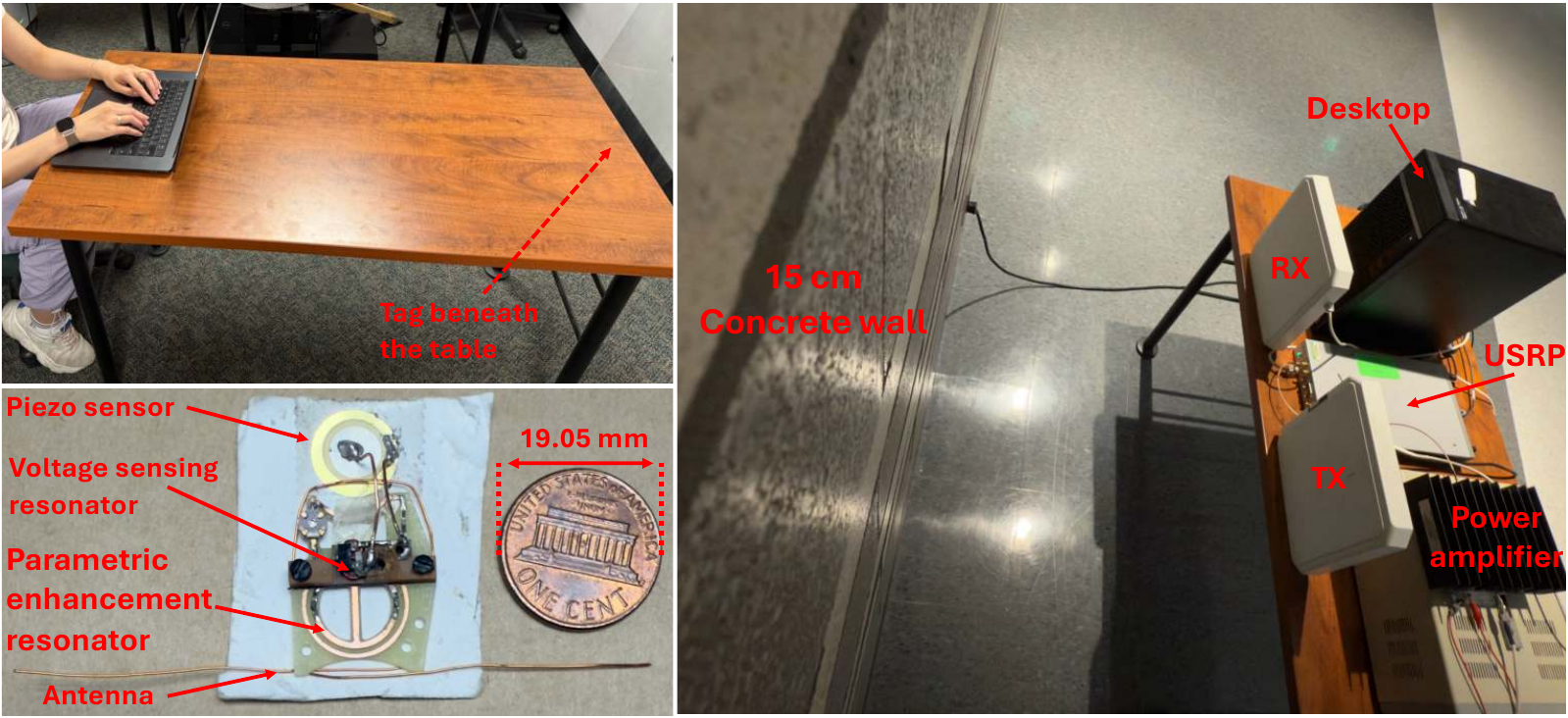}
    \caption{
    Experimental settings. 
    }
    \label{fig:exp_setting}
\end{figure}

\textbf{RF Tag.}
Fig. \ref{fig:exp_setting} illustrates our fabricated tag, comprising a piezoelectric sensor, a VSR, and a PER.
The PER features a circular inductor etched on a 0.8 mm thick G10  fiberglass-epoxy substrate, with an inner diameter of 13.5 mm and an outer diameter of 14.5 mm. Its upper/right and lower/left semi-circular sections are interrupted by split gaps, each filled with varactor diodes providing a capacitance of 9.1~pF. This configuration supports a \textit{circular} resonance mode. A horizontal conductor bridges the two virtual voltage nulls of this mode, enabling a \textit{butterfly} resonance mode.

The VSR is constructed by winding 32-gauge enameled copper wire around two rods, each 1.5 mm in diameter and spaced 1.8 mm apart. The coil consists of five counterclockwise turns on the first rod, followed by a single clockwise turn on the second. The two wire ends are connected to a pair of head-to-head varactor diodes with a capacitance of 3 pF. The piezoelectric transducer connects one sensing electrode to the common cathode and the other to the common anode.


\textbf{RF Reader.}
Fig.~\ref{fig:exp_setting} also shows our RF reader, which consists of a USRP N310, a power amplifier (PA), two directional antennas, and a laptop.
The RF reader transmits a 30~dBm excitation signal at 915~MHz. 
The tag generates backscattered signals at 515~MHz.
All signal processing is performed on the laptop using Python.
For LLM-guided online adaptation, we deploy a locally hosted \texttt{LLaMA-3.2-3B} model \cite{dubey2024llama}, which interfaces with the keystroke classifier to provide online supervision.

\subsection{Experimental Setup and Performance Metrics}



\begin{figure}[t]
    \centering
    \includegraphics[width=0.99\linewidth]{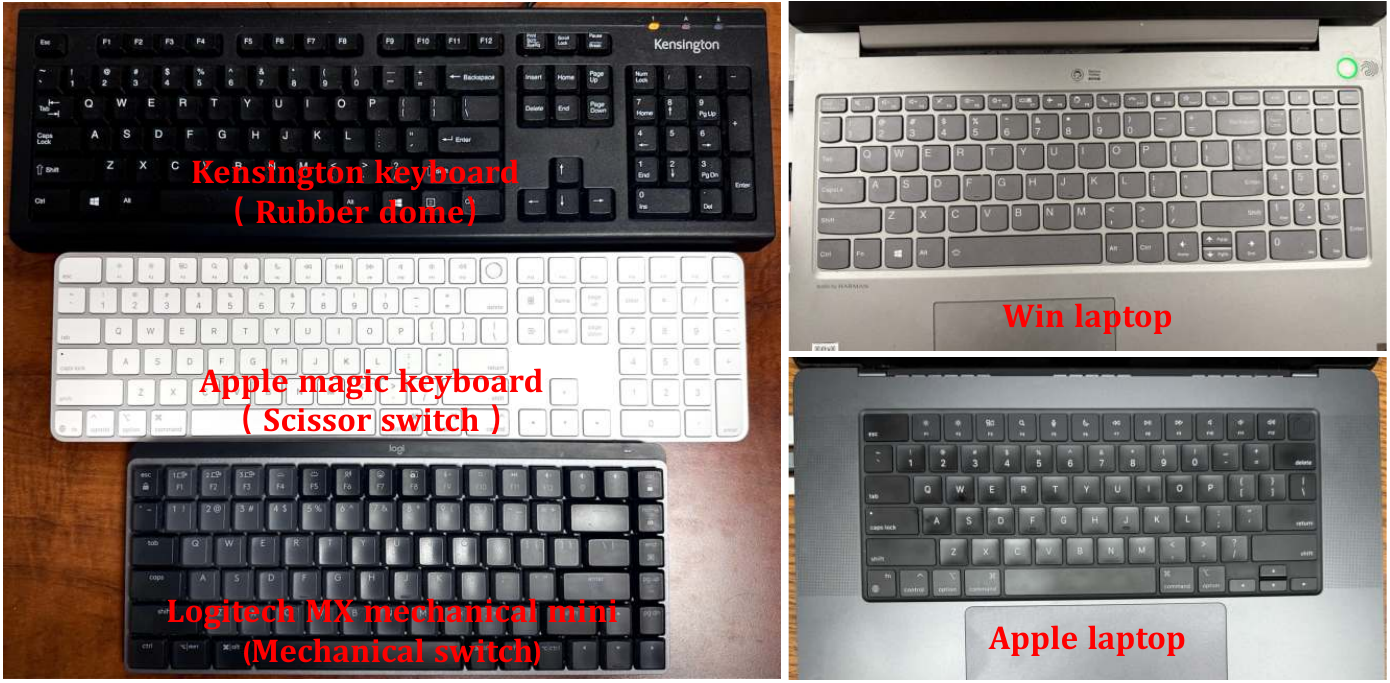}
    \caption{
    Keyboards and laptops.
    }
    \label{fig:keyboard_pic}
\end{figure}

\textbf{Experimental Setup.}
All experiments were conducted in an indoor office environment using five representative input devices: a mechanical keyboard, a scissor-switch keyboard, a rubber dome keyboard, a Windows laptop, and an Apple laptop, as shown in Fig.~\ref{fig:keyboard_pic}.
Unless otherwise noted, the experiments were performed on a wooden table; the backscatter tag was affixed to its underside; the horizontal distance between tag and keyboard was about 0.9~m; and the RF reader was positioned 3~m from the tag.

Our experiments involved 10 participants with varied typing styles and speeds. 
Every participant was asked to type both predefined and freeform inputs. 
The input corpus was designed to ensure full keyboard coverage, including
all alphabetic keys (`a'–`z'),
numeric digits (`0'–`9'),
control keys (e.g., space, enter, and delete/backspace),
common punctuation marks (e.g., `.', `,', `?', and `!'),
and 
special characters and symbols (e.g., \texttt{@}, \texttt{\$}, \texttt{\%}, and \texttt{\&}).

We consider two classes of typing content with distinct semantic structures in our experiments.
\textit{(i) Natural Language Input:} 
These are sentences and paragraphs from general-purpose corpora (e.g., Wikipedia articles, email replies, and news excerpts). 
This type of content reflects conversational or document-style typing where LLM's linguistic priors and contextual reasoning are most effective. 
\textit{(ii) Non-linguistic Input:} 
This type of content includes random character sequences, passwords, email addresses, and command-line strings that lack grammatical structure or semantic context. 


\begin{figure*}[!t]
    \centering

    \begin{subfigure}[t]{0.31\textwidth}
        \centering
        \includegraphics[width=\linewidth]{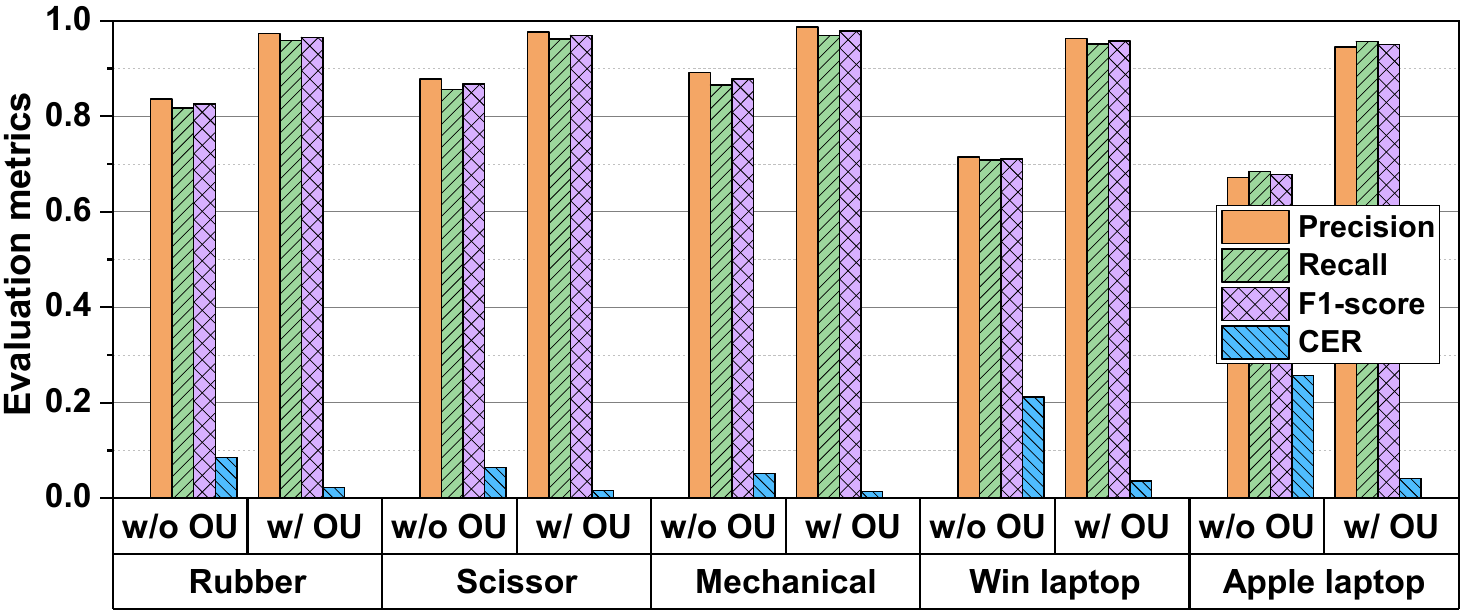}
        \caption{Performance for different keyboards.}
        \label{fig:keyboards}
    \end{subfigure}
    \hfill
    \begin{subfigure}[t]{0.31\textwidth}
        \centering
        \includegraphics[width=\linewidth]{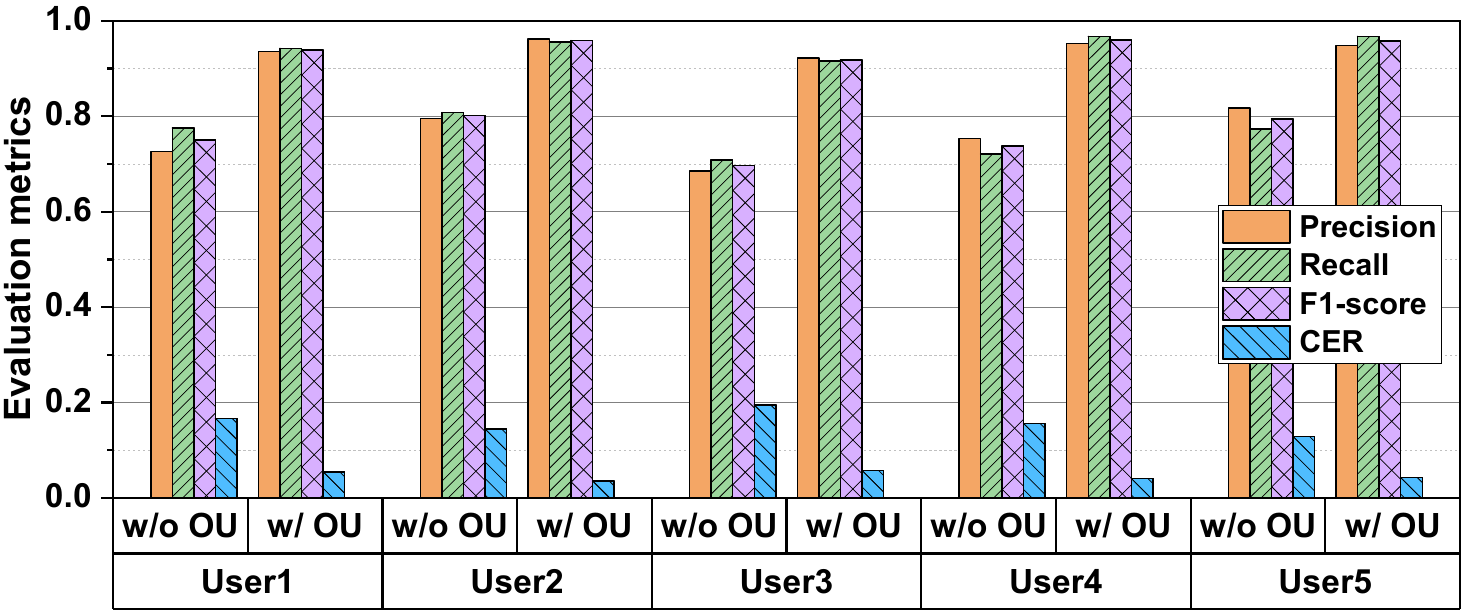}
        \caption{Performance across different users.}
        \label{fig:users}
    \end{subfigure}
    \hfill
    \begin{subfigure}[t]{0.33\textwidth}
        \centering
        \includegraphics[width=0.925\linewidth]{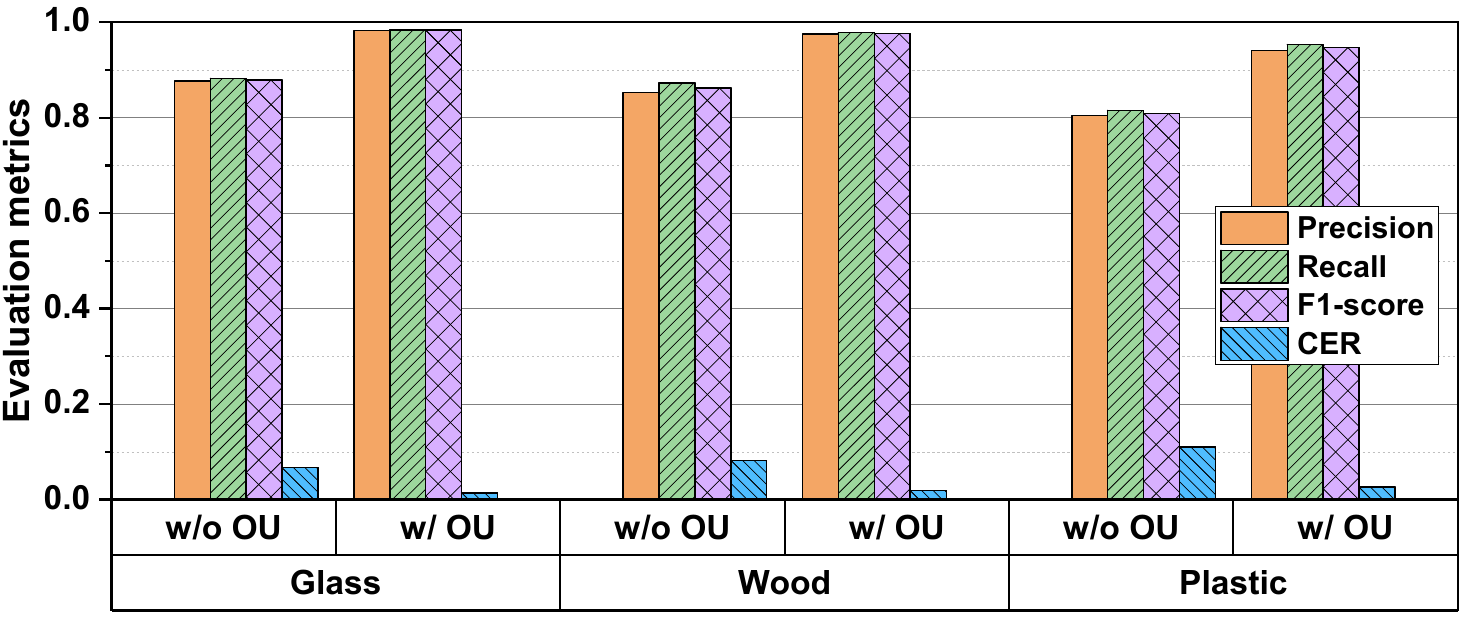}
        \caption{Performance with different table surfaces.}
        \label{fig:surfaces}
    \end{subfigure}
    \caption{Keystroke detection performance under different conditions.  [\textbf{OU}: LLM-guided \textbf{O}nline \textbf{U}pdating.]}
    \label{fig:keystroke_conditions}
\end{figure*}

\textbf{Performance Metrics.}
We evaluated the performance of \pname using four standard metrics.
\textit{(i) Precision:} The ratio of correctly-predicted keystrokes to all predicted keystrokes.
\textit{(ii) Recall:} The ratio of correctly-detected keystrokes to all ground-truth keystrokes.
\textit{(iii) F1-Score:} The harmonic mean of precision and recall, reflecting the balance between them.
\textit{(iv) Character Error Rate (CER):} the normalized Levenshtein distance between the predicted string and the ground truth, defined as:
$\text{CER} = \frac{S + D + I}{N}$, 
where $S$, $D$, and $I$ denote the number of substitutions, deletions, and insertions, respectively, and $N$ is the total number of characters in the ground-truth string.


\subsection{Main Results}
Fig.~\ref{fig:keystroke_conditions} reports the overall performance of \pname across three practical dimensions: keyboard type, user variability, and supporting surface. Overall, \pname achieves strong performance, exceeding 90\% detection precision across all conditions.
For each setting, we compare the performance of \pname with and without LLM-guided online adaptation. The results show that LLM-guided adaptation provides substantial improvements.
It typically boosts the detection accuracy by more than 10\%. This demonstrates the effectiveness of incorporating LLM-guided contextual correction and online refinement into our design.
In the following, we present the detailed results for each condition.

\textbf{Linguistic versus Non-linguistic keystroke Inputs.}

To understand how semantic structure affects \pname, we evaluate the system on both linguistic and non-linguistic typing inputs. Table~\ref{tab:ling_nonling} summarizes the experimental results.
Without online updating, \pname achieves 84.8\% accuracy on linguistic inputs and 80.1\% on non-linguistic inputs. With LLM-guided online adaptation, the accuracy improves to 97.7\% and 88.3\%, respectively. The larger gain for linguistic inputs is expected, as natural-language context provides strong semantic constraints that help correct classification errors.

The improvement on non-linguistic inputs arises from the mixed-input nature of real typing streams. In our dataset, linguistic entries constitute the majority (86.7\%), while non-linguistic entries account for the minority (13.3\%). When users type linguistic content, the LLM-guided online adaptation module generates pseudo-labels to refine the keystroke classifier, enabling it to adapt to the user’s typing patterns and keyboard characteristics. The refined classifier then yields more accurate predictions for subsequent non-linguistic inputs as well.


\input{tables/linguistic}

\textbf{Impact of Keyboard Types.}
We evaluate \pname across the five representative keyboard types as shown in Fig.~\ref{fig:keyboard_pic}.  
Fig.~\ref{fig:keyboards} presents our measurement results. 
External keyboards generate stronger and more consistent vibrations due to longer key travel and higher actuation force, resulting in clearer signal imprints and higher detection accuracy. In contrast, the Apple laptop exhibits the lowest detection accuracy.
This can be attributed to its short key travel distance, low-profile mechanism, and rigid chassis, all of which attenuate vibrations and reduce SNR at the tag.  
As shown in Fig.~\ref{fig:keyboards}, LLM-guided online adaptation enables \pname to achieve consistently high accuracy across all keyboard types, demonstrating strong robustness to the variance of keyboard types.

\textbf{Performance Across Different Users.}
Generalizing across users is a key challenge, as typing style, finger force, posture, and keypress dynamics vary substantially between individuals. Even when using identical keyboards, subtle factors such as key aging, manufacturing tolerances, and acoustic differences introduce user-specific variations in the signal.  
To assess robustness, we conduct cross-user evaluations in which the model is trained on one user group and tested on entirely unseen users.  
Fig.~\ref{fig:users} presents our results. 
The performance without LLM-guided adaptation degrades noticeably due to user-dependent signal shifts. In contrast, \pname maintains high accuracy through its LLM component, which enforces linguistic consistency and provides pseudo-label supervision for online refinement. This allows the classifier to adapt to new users, yielding strong generalization without requiring user-specific training data.

\textbf{Impact of Keyboard-Supporting Surfaces.}
We further evaluate performance across three common surface materials on which the keyboard and tag may be placed: glass, wood, and plastic.  
Fig.~\ref{fig:surfaces} reports our results.
The non-adaptive model performs best on glass due to its rigid and uniform structure, which preserves vibrations accurately. Wood provides moderate vibration transmission and yields comparable accuracy. Plastic performs the worst because its higher damping and irregular internal structure attenuate vibrations more significantly.  
Despite these disparities, \pname achieves uniformly high performance across all surfaces. The combination of dual-path sensing and LLM-guided online adaptation mitigates surface-induced variability, enabling stable performance in diverse real-world environments.

\input{tables/eva_distance}

\subsection{Impact of Tag Distance and Orientation}

\textbf{Impact of Tag-to-Keyboard Distance.}
To assess how proximity between the keyboard and the RF tag affects keystroke inference, we vary their horizontal distance from 10~cm to 120~cm. This range captures both tightly coupled and loosely coupled setups, reflecting practical considerations such as space constraints or concealment requirements.

Table~\ref{tab:tag_dis} reports the experimental results. As expected, performance degrades gradually as the distance increases due to vibration attenuation and weakened mechanical coupling. Larger distances reduce the SNR at the tag, making keystroke-induced vibrations more susceptible to noise. This effect becomes particularly evident beyond 100~cm, where the character error rate rises sharply without LLM-guided adaptation.
However, with LLM-guided online adaptation, \pname maintains high accuracy across all tested distances. The LLM leverages linguistic context to refine predictions and compensate for distance-induced signal degradation. These results demonstrate that \pname remains effective even when the keyboard is placed more than one meter away from the tag, enabling flexible and covert deployment.

\textbf{Impact of Tag-to-Reader Distance.}
Another critical factor for the key-typing eavesdropping attack is the tag-to-reader distance, which determines the effective reading range of the RF system. We evaluate this by placing the RF reader at distances ranging from 1~m to 8~m. Beyond 6~m, line-of-sight (LoS) is obstructed by furniture, representing realistic indoor obstructions.
Table~\ref{tab:tag_dis1} shows the results. \pname consistently achieves high accuracy across all distances, maintaining a 90.5\% F1-score and only a 5.1\% character error rate at 8~m. In contrast, removing the LLM component leads to significant degradation beyond 5~m, where lower SNR and increased distortion cause the error rate to rise sharply. These findings highlight the importance of combining physical-layer robustness with semantic-level correction to enable long-range keystroke inference. Overall, \pname remains effective even under remote or partially obstructed surveillance conditions.

\input{tables/eva_distance1}

\begin{figure}
    \centering
    \includegraphics[width=0.99\linewidth]{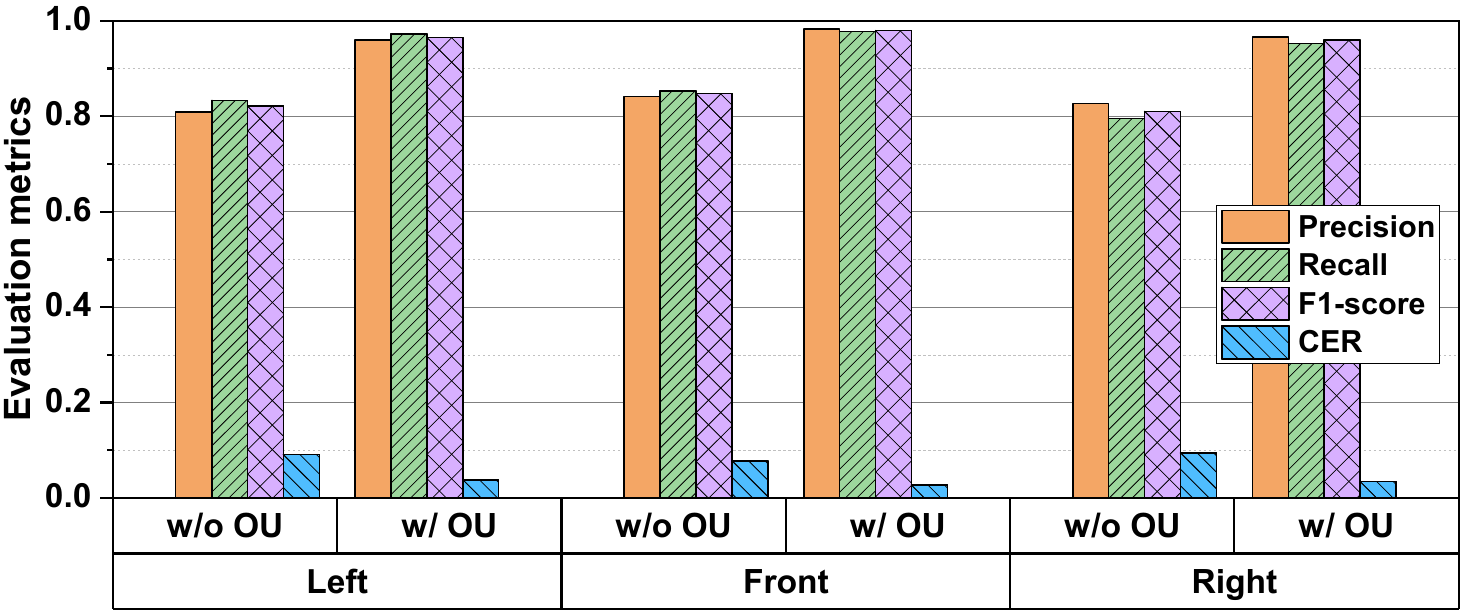}
    \caption{
        Impact of the RF tag's placement. [\textbf{OU}: LLM-guided \textbf{O}nline \textbf{U}pdating.]
    }
    \label{fig:placement}
\end{figure}




\textbf{Impact of Tag Orientation.}
We further evaluate the performance of \pname when the tag is positioned in three orientations relative to the keyboard: facing the front, left, or right. 
Fig.~\ref{fig:placement} shows the experimental results. 
\pname maintains strong performance across all orientations when equipped with LLM-guided adaptation.
Among the three, the front-facing orientation yields the highest accuracy due to more symmetrical vibration propagation along the table surface. Lateral placements (left or right) experience asymmetric energy distribution and higher damping, resulting in weaker received signals and slightly lower accuracy. Nonetheless, \pname remains robust across all configurations, demonstrating resilience to variations in tag placement.


\begin{table}[t]
    \centering
    \captionof{table}{Ablation study.}
    \label{tab:ablation1}
    \resizebox{\linewidth}{!}{
    \begin{tabular}{l|cccc}
    \hline
    Configuration               & \multicolumn{4}{l}{\ \ \ \ \ \ \ \ \ \ \ \ \ \ \ \ \ \ \ \ Choice} \\ \hline
    Coarse-Grained Features                 & \ding{52}     & \ding{52}    & \ding{52}    & \ding{52}    \\
    Fine-Grained Features              & \ding{56}     & \ding{52}    & \ding{52}    & \ding{52}    \\
     LLM (Post-Processing)             & \ding{56}     & \ding{56}    & \ding{52}    & \ding{56}    \\
    LLM (Online Updating)        & \ding{56}     & \ding{56}    & \ding{56}    & \ding{52}    \\ \hline
    Precision  &   64.7\%   &  82.6\%    &   87.9\%   &  98.1\%    \\
    Recall     &   63.1\%    &   82.3\%   &   85.5\%   &    97.6\%  \\
    F1-Score   &   63.9\%    &   82.4\%   &   86.7\%   &    97.8\%  \\
    Character Error Rate        &    19.6\%   &  10.8\%    &   6.7\%   &     1.5\%
    \\\hline
    \end{tabular}
    }
\end{table}

\subsection{Ablation Study}

To quantify the contribution of each component in the RF reader system, we conduct an ablation study by selectively enabling or disabling key modules. Our experiments evaluate four configurations:
\textit{(i) Coarse baseline:} This baseline uses only coarse-grained features extracted directly from raw RF signals. A CoAtNet \cite{dai2021coatnet} classifier is trained without hierarchical fusion or language-model assistance.
\textit{(ii) Fine-grained features:} We enhance the baseline with fine-grained representations extracted via a high-resolution branch, fused with coarse features using a cross-attention module.
\textit{(iii) LLM post-processing:} An LLM-guided decoder is added for context-aware refinement, correcting classification outputs using language semantics.
\textit{(iv) LLM in the training loop:} Finally, we integrate the LLM into the training loop through a feedback mechanism that penalizes semantically inconsistent predictions. This is the final RF reader used in \pname. 

Table~\ref{tab:ablation1} summarizes the results, demonstrating that each component provides a meaningful performance gain. Incorporating fine-grained features improves the F1-score from 63.9\% to 82.4\%, and LLM post-processing further increases it to 86.7\%. Enabling LLM-guided online adaptation yields the best performance, boosting the F1-score to 97.8\% and reducing the character error rate to 1.5\%.

\begin{figure}[t]
    \centering
    \includegraphics[width=0.95\linewidth]{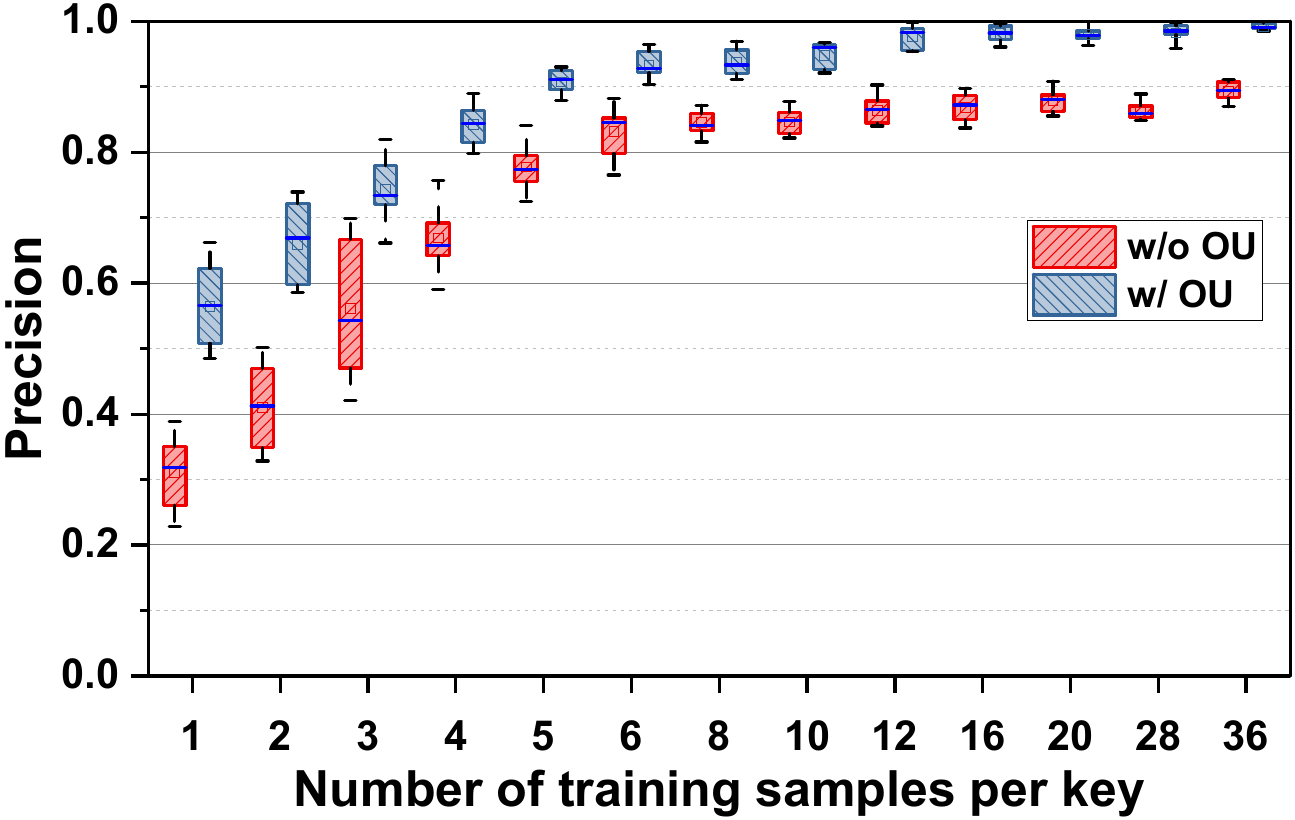}
    \caption{
        Impact of pre-training dataset size on the performance of \pname. 
        [\textbf{OU}: LLM-guided \textbf{O}nline \textbf{U}pdating.]
    }
    \label{fig:samples}
\end{figure}%

\begin{figure}[t]
    \centering
    \includegraphics[width=0.95\linewidth]{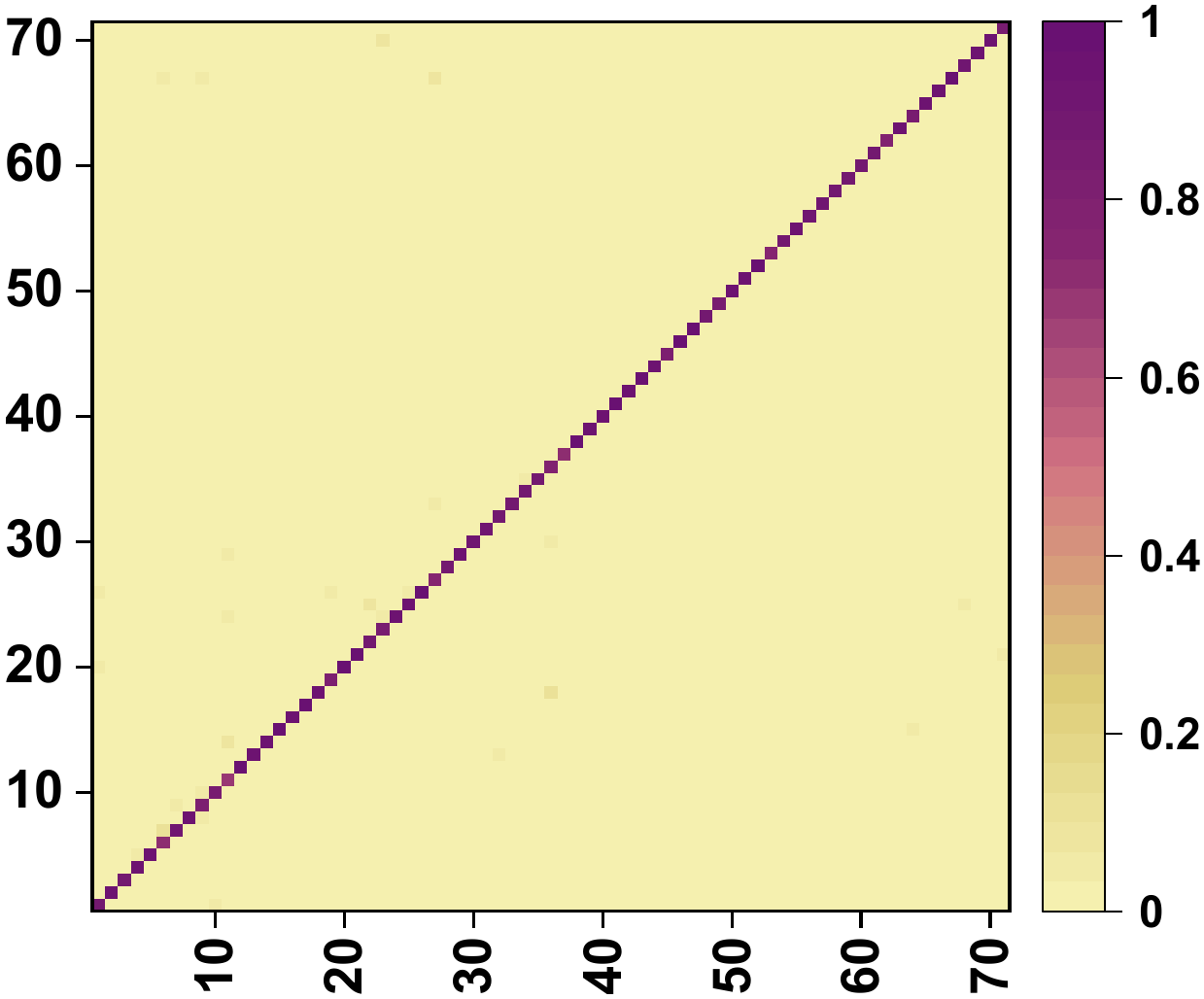}
    \caption{Through-wall keystroke detection accuracy.}
    \label{fig:heatmap}
\end{figure}

\subsection{Impact of Pre-training Dataset Size}

To assess how pre-training dataset size influences \pname's performance when training data comes exclusively from non-target users and keyboards, we vary the number of samples per key from 1 to 36 and evaluate the model on the target test set.
Fig.~\ref{fig:samples} presents our experimental results.
It can be seen that increasing the number of samples consistently improves performance without LLM adaptation, but the gain plateaus beyond 20 samples per key, indicating diminishing returns under non-target-only training. In contrast, with LLM-guided online adaptation, \pname achieves high accuracy with as few as 4--8 samples per key. The LLM facilitates effective domain adaptation by injecting linguistic priors into the classifier, even in the absence of victim-specific data. These results demonstrate that \pname is robust in low-data regimes and can be deployed without requiring any target-specific training data.


\subsection{Through-Wall Keystroke Detection}

To evaluate the covert sensing capability of \pname, we examine its performance in through-wall scenarios where the RF reader is placed outside the room, while the victim and passive tag remain inside, separated by a 15~cm concrete wall. This setup creates a fully non-line-of-sight condition, with the direct RF path completely blocked. Unlike long-distance attenuation, through-wall propagation introduces additional challenges such as multipath distortion and dielectric-induced phase shifts, which degrade signal quality and complicate demodulation.

Fig.~\ref{fig:heatmap} presents our experimental results, where both x-axis and y-axis represent keystroke indices (`a'-`z', `A'-`Z', `0'-`9', and special symbols). \pname maintains a high inference accuracy even under this challenging condition. This robustness can be attributed to:
(i) the inherent resilience of the dual-resonator tag’s frequency modulation to noise and fading;
(ii) dual-path sensing that captures both solid-borne and air-borne signal components, mitigating blockage effects; and
(iii) LLM-guided online adaptation that corrects channel-induced distortions.

%% file: tables/linguistic.tex
\begin{table}[t]
\centering
\caption{Performance of \pname when the typing input is linguistic and non-linguistic text.}
\label{tab:ling_nonling}
\resizebox{0.48\textwidth}{!}{%

\begin{tabular}{l|c|c}
\hline
\textbf{Input Type} & \textbf{W/o Online Updating} & \textbf{W/ Online Updating}\\
\hline
Linguistic input & 84.8\% & 97.7\%  \\
Non-linguistic input & 80.1\% & 88.3\%  \\
\hline
\end{tabular}
}
\end{table}

%% file: tables/eva_distance.tex
\begin{table}
\centering
\scriptsize
\caption{Keystroke detection performance at different tag-to-keyboard distances. [\textbf{OU}: LLM-guided \textbf{O}nline \textbf{U}pdating.]}
\label{tab:tag_dis}
\resizebox{\linewidth}{!}{%
\begin{tabular}{c|cc|cc|cc|cc} 
\hline
\multirow{2}{*}{} & \multicolumn{2}{c|}{Precision} & \multicolumn{2}{c|}{Recall} & \multicolumn{2}{c|}{F1-score} & \multicolumn{2}{c}{CER} \\ \cline{2-9} 
& \!\!\!\!w/o OU   \!\!\!\!\!\!\!\!     & \!\!\!\!w/ OU   \!\!\!\!\!\!\!\!    & \!\!\!\!w/o OU    \!\!\!\!\!\!\!\!  & \!\!\!\!w/ OU  \!\!\!\!\!\!\!\!    & \!\!\!\!w/o OU    \!\!\!\!\!\!\!\!   & \!\!\!\!w/ OU   \!\!\!\!\!\!\!\!   & \!\!\!\!w/o OU   \!\!\!\!\!\!\!\!    & \!\!\!\!w/ OU  \!\!\!\!\!\!\!\!  \\ \hline
\!\!\!\!10 cm\!\!\!\!             & 90.5\%	&98.5\% 	&88.2\%	 &97.9\% 	&89.3\% 	&98.2\% 	&5.1\%	 &1.3\%      \\ \hline
\!\!\!\!20 cm\!\!\!\!            &91.3\%	&98.1\%  	&87.8\%  	&97.4\%  	&89.5\%  	&97.7\%  	&4.9\%  	&1.5\%             \\ \hline
\!\!\!\!30 cm\!\!\!\!            &90.1\%  	  &97.3\%  	  &89.5\%  	  &98.1\%  	 &89.8\%  	&97.7\%  	&5.4\%  	&2.1\%           \\ \hline
\!\!\!\!40 cm\!\!\!\!            &89.6\%  	 &98.3\%  	 &89.1\%  	 &97.5\%  	 &89.3\%  	 &97.9\%  	 &5.5\%  	 &1.7\%            \\ \hline
\!\!\!\!50 cm\!\!\!\!            & 87.3\%  	 & 98.2\%  	 &86.7\%  	    &97.3\%  	    &87.0\%  	    &97.7\%  	   &6.4\%      &1.6\%           \\ \hline
\!\!\!\!60 cm\!\!\!\!            &88.7\%  	    &97.4\%  	    &88.1\%  	    &96.8\%  	    &88.4\%  	    &97.1\%  	    &6.2\%  	    &1.9\%            \\ \hline
\!\!\!\!70 cm\!\!\!\!            & 87.1\%  	    &97.6\%  	    &85.7\%  	    &96.5\%  	    &86.4\%  	    &97.0\%  	    &6.9\%  	    &1.8\%            \\ \hline
\!\!\!\!80 cm\!\!\!\!            & 86.5\%  	    &97.8\%  	    &87.1\%  	    &97.4\%  	    &86.8\%  	    &97.6\%  	    &7.1\%  	    &1.9\%           \\ \hline
\!\!\!\!90 cm\!\!\!\!            & 85.8\%  	   &97.5\%  	   &85.7\%  	   &97.8\%  	   &85.7\%  	   &97.6\%  	   &7.5\%  	   &2.1\%         \\ \hline
\!\!\!\!100 cm\!\!\!\!            &82.2\%  	    &96.7\%  	    &83.6\%  	    &96.5\%  	    &82.9\%  	    &96.6\%  	    &8.2\%  	    &2.3\%         \\ \hline
\!\!\!\!110 cm\!\!\!\!            & 82.7\%  	    &95.4\%  	    &82.3\%  	    &95.1\%  	    &82.5\%  	    &95.2\%  	    &8.4\%  	    &2.8\%         \\ \hline
\!\!\!\!120 cm\!\!\!\!            & 81.5\%  	      &95.5\%  	      &80.3\%  	      &95.7\%  	      &80.9\%  	      &95.6\%  	      &9.3\%  	      &2.7\%         \\ \hline
\end{tabular}
}
\end{table}

%% file: tables/eva_distance1.tex
\begin{table}
\centering
\scriptsize
\caption{Keystroke detection performance at different tag-to-reader distances. [\textbf{OU}: LLM-guided \textbf{O}nline \textbf{U}pdating.]}
\label{tab:tag_dis1}
\resizebox{\linewidth}{!}{%
\begin{tabular}{c|cc|cc|cc|cc} 
\hline
\multirow{2}{*}{} & \multicolumn{2}{c|}{Precision} & \multicolumn{2}{c|}{Recall} & \multicolumn{2}{c|}{F1-score} & \multicolumn{2}{c}{CER} \\ \cline{2-9} 
& \!\!\!\!w/o OU   \!\!\!\!\!\!\!\!     & \!\!\!\!w/ OU   \!\!\!\!\!\!\!\!    & \!\!\!\!w/o OU    \!\!\!\!\!\!\!\!  & \!\!\!\!w/ OU  \!\!\!\!\!\!\!\!    & \!\!\!\!w/o OU    \!\!\!\!\!\!\!\!   & \!\!\!\!w/ OU   \!\!\!\!\!\!\!\!   & \!\!\!\!w/o OU   \!\!\!\!\!\!\!\!    & \!\!\!\!w/ OU  \!\!\!\!\!\!\!\!  \\ \hline
\!\!\!\!1 m\!\!\!\!             & 88.5\%  	&98.3\%  	&87.8\%  	&98.2\%  	&88.1\%  	&98.3\%  	&5.9\%  	&1.2\%        \\ \hline
\!\!\!\!2 m\!\!\!\!            &86.5\%   	&97.7\%  	&86.2\%  	&97.9\%  	&86.4\%  	&97.8\%  	&7.1\%  	&1.9\%             \\ \hline
\!\!\!\!3 m\!\!\!\!            &86.1\%  	  &97.8\%  	  &85.3\%  	  &97.1\%  	 &85.7\%  	&97.4\%  	&7.4\%  	&2.0\%           \\ \hline
\!\!\!\!4 m\!\!\!\!            &83.5\%  	 &96.3\%  	 &82.1\%  	 &95.8\%  	 &82.8\%  	 &96.0\%  	 &8.8\%  	 &2.8\%            \\ \hline
\!\!\!\!5 m\!\!\!\!            & 82.8\%  	 & 96.1\%  	 &82.5\%  	    &96.0\%  	    &82.7\%  	    &96.0\%  	   &9.1\%      &2.9\%           \\ \hline
\!\!\!\!6 m\!\!\!\!            &78.7\%  	    &94.8\%  	    &77.8\%  	    &95.1\%  	    &78.2\%  	    &95.0\%  	    &14.6\%  	    &3.1\%            \\ \hline
\!\!\!\!7 m\!\!\!\!            & 70.5\%  	    &91.7\%  	    &71.3\%  	    &91.5\%  	    &70.9\%  	    &91.6\%  	    &20.1\%  	    &4.9\%           \\ \hline
\!\!\!\!8 m\!\!\!\!            & 66.3\%  	      &90.7\%  	      &67.8\%  	      &90.3\%  	      &67.0\%  	      &90.5\%  	      &23.7\%  	      &5.1\%         \\ \hline
\end{tabular}
}
\end{table}

%% file: 8_defence_against_keystroking_attack.tex
As keystroke inference attacks via passive RF backscatter become increasingly practical and covert, defending against such threats is critical to safeguarding user privacy. In this section, we discuss potential defense strategies across four levels: physical-layer obfuscation, signal-space mitigation, semantic-level hardening, and proactive defense.

\textbf{Physical and Environmental Obfuscation.}
One natural line of defense is to disrupt the physical transduction pathway between keystrokes and RF signal perturbations. Since our system relies on subtle mechanical and acoustic vibrations transmitted through surfaces, introducing damping materials (e.g., acoustic foam, rubber mats, or vibration-absorbing desk layers) can significantly attenuate the signal before it reaches the backscatter tag.

\textbf{Signal-Level Countermeasures.}
From the perspective of RF signal processing, defenders can employ \textit{signal obfuscation} techniques such as artificial vibration injection or ambient RF noise generation. For instance, actuators placed near the tag could emit controlled low-level vibrations during typing events to corrupt keystroke-induced patterns.
Moreover, defenders could deploy \textit{RF fingerprinting traps}, which monitor the backscattered signal and flag suspicious passive tags based on abnormal reflection characteristics or unexpected resonance patterns.

\textbf{Semantic and Language-Level Defenses.}
Given that our system leverages LLM to refine noisy outputs using linguistic priors, a promising direction is to \textit{minimize the utility of such priors}. One way is to introduce randomness into typing behavior, such as inserting decoy keystrokes or using randomized input sequences, which disrupts regularity and hinders language modeling~\cite{taheritajar2024survey,zhuang2009keyboard}.
Users can also adopt \textit{linguistic-resistant input schemes} when entering sensitive information (e.g., using password managers or on-screen keyboards that randomize key positions). Since these schemes break natural language syntax, the LLM-guided correction becomes less effective, especially when domain adaptation is not feasible.

\textbf{Toward Proactive Defense Frameworks.}
Long-term defense against RF-based keystroke attacks must be embedded into \textit{multi-layered system architectures}. This includes (i) trusted computing modules that detect unauthorized sensing hardware, (ii) sensor access control policies that flag anomalous passive tags, and (iii) adversarial learning models that identify and degrade LLM-guided inference in hostile environments~\cite{han2025rf, yu2021security}.
Our attack shows that high-accuracy keystroke inference is feasible using passive RF signals and LLM supervision, underscoring the need for defenses that anticipate both new sensing modalities and intelligent post-processing models in adversarial pipelines.

%% file: 2_related_work.tex
\input{tables/comparison}
Keystroke eavesdropping attacks can generally be classified based on their sensing modalities, sensor placements, training requirements, and generalizability
\cite{han2025rf, sikder2021survey, yu2021security, taheritajar2024survey}.
Table~\ref{tab:related_work} summarizes existing keystroke eavesdropping work.
We detail them in the following two categories. 

\subsection{Microphone-Based Keystroke Eavesdropping}

Research has been conducted over decades to study keystroke eavesdropping using microphones.
Asonov et al. \cite{asonov2004keyboard} first demonstrated that acoustic signals from keyboards differ among keys, achieving a 79\% detection accuracy rate at 1 m. However, their method requires extensive labeled data and lacks robustness to environmental variability.

To address these limitations, Zhuang et al. \cite{zhuang2009keyboard} introduced an unsupervised method that utilized cepstrum features, Hidden Markov Models (HMM), and iterative feedback-based incremental learning. 
Their approach does not need labeled data and achieved over 96\% character-level accuracy.
%
Zhu et al. \cite{zhu2014context} further improved unsupervised acoustic attacks by introducing a context-free model using microphones placed 25 cm away. 
Their method achieved 72.2\% accuracy without labeled training data.

In contrast, Berger et al. \cite{berger2006dictionary} introduced dictionary-constrained attacks, leveraging known language models to map acoustic signals to likely words. 
It claimed a 90\% detection accuracy. 
Some work \cite{compagno2017don,cecconello2019skype} has investigated keystrokes' acoustic leakage in VoIP applications. 
These studies demonstrated that keystroke sounds remain distinguishable through compressed audio channels, achieving character detection accuracies exceeding 90\%.
Later efforts enhanced the robustness of this approach under realistic settings. Bai et al. \cite{bai2021know} achieved 91.5\% accuracy in noisy environments using refined inference techniques. 
Giallanza et al. \cite{giallanza2019keyboard} adopted DNN ensembles on audio captured by smartphone arrays, achieving 70.6\% accuracy.

KeystrokeSniffer \cite{huang2024keystrokesniffer} demonstrated inference from audio recorded 15\,cm away using commodity smartphones, achieving 79.5\% accuracy. Auditory Eyesight \cite{tu2023auditory} utilized microsecond-level timing to reach 90.8\% accuracy at 50~cm. Liu et al. \cite{liu2015snooping} further pushed the limits by using millimeter-scale acoustic ranging to reach 97.7\% accuracy with minimal training and only one microphone.
Slater et al. \cite{slater2019robust} introduced an end-to-end deep learning pipeline for keystroke inference under overlapping waveforms and fast typing.
%
Wearable-based approaches, such as SIA \cite{meteriz2021sia}, exploited smartwatch microphones to maintain discretion and achieved 85\% recognition accuracy.

While these microphone-based works demonstrated the feasibility of acoustic side-channel attacks, their assumptions and training requirements remain questionable in realistic settings. 
Some require extensive supervised training data \cite{asonov2004keyboard,harrison2023practical}, task- or setup-specific modeling \cite{slater2019robust,bai2021know}, or carefully controlled recording conditions \cite{asonov2004keyboard,liu2015snooping,harrison2023practical}; 
others reduce deployment-specific training by relying on strong prior knowledge such as dictionary or language constraints \cite{berger2006dictionary,zhuang2009keyboard}.
In general, their performance often degrades across different keyboards, users, and environments, limiting robust generalization in real deployments.
\pname addresses these limitations by replacing conspicuous nearby microphones with a covert RF backscatter tag, enabling through-wall operation, and combining dual-path feature design with LLM-guided online adaptation to achieve more robust keystroke inference across diverse real-world deployments.

\subsection{Keystroke Eavesdropping Using Other Sensors}

Alternative sensor modalities have also been explored in different ways for keystroke inference. Marquardt et al. \cite{marquardt2011sp} utilized accelerometers to sense keyboard vibrations, achieving 80\% accuracy at close range. 
OverHear \cite{wijewickrama2023overhear} enhanced inference accuracy by combining accelerometer and microphone data, effectively maintaining stealth.
Wireless sensing methods have also proven feasible for this attack. 
WiKey \cite{ali2015keystroke} and Chen et al. \cite{chen2015tracking} leveraged WiFi Channel State Information (CSI) and custom RF front-ends, respectively, achieving high accuracy (90-96\%) but requiring complex multi-antenna setups. 
Fang et al. \cite{fang2018no} developed a training-free CSI-based inference approach, enhancing deployment flexibility while achieving 95.3\% accuracy.
These alternative approaches demonstrate that keystroke leakage is not confined solely to acoustic channels. Methods exploiting vibrations and wireless signals often match or exceed acoustic methods under ideal conditions but generally require specialized hardware setups and calibration, limiting their practical generalization and stealth.

%% file: tables/comparison.tex
\begin{table*}[!t]
\centering
\caption{
Overview and comparison of keystroke eavesdropping attack methodologies.
Key to understanding this comparison:
\textbf{Non-Invasive:} Attack does not require visible or suspicious sensor placement near the victim's device.
\textbf{Closed Vocabulary:} Method is restricted to recognizing only a predefined, finite set of words or characters.
\textbf{Symbols:} \CIRCLE~High; \LEFTcircle~Moderate; \Circle~Low generalization/performance. -=Unknown/Not Provided.
}
\resizebox{\textwidth}{!}{%
\begin{tabular}{c|c|c|c|c|c|c|c|c|c|c}
\hline
\multirow{2}{*}{\textbf{Previous Work}} & \multirow{2}{*}{\textbf{Sensor} \& \textbf{Signal Type}} & \multirow{2}{*}{\!\!\!\makecell{\textbf{Attack}\\ \textbf{Distance}}\!\!\!} & \multirow{2}{*}{\!\!\!\makecell{\textbf{Non-} \\ \textbf{Invasive}}\!\!\!} & \multirow{2}{*}{\!\!\!\!\makecell{\textbf{Through}\\ \textbf{Wall}}\!\!\!\!} & \multirow{2}{*}{\makecell{\textbf{Training/ML} \\ \textbf{Dependency}}} & \multirow{2}{*}{\!\!\!\makecell{\textbf{Closed}\\ \textbf{Vocabulary?}}\!\!\!} & \multirow{2}{*}{\textbf{Target key Set}} & \multicolumn{2}{c|}{\!\!\!\textbf{Generalization ability}\!\!\!} & \multirow{2}{*}{\!\!\!\textbf{Accuracy}\!\!\!} \\ \cline{9-10}
 &  &  &  &  &  &  &  & User & \!\!\!\!Environment\!\!\!\! &  \\ \hline
 
Zhu \textit{et al}. \cite{zhu2014context} & Smartphone microphones, Acoustic & 25 cm & \xmark & \xmark & \xmark\;Train, \xmark\;ML & \xmark & `a-z' + 3 Special keys & \CIRCLE & \LEFTcircle & 72.2\% \\ \hline

Liu \textit{et al}. \cite{liu2015snooping} & Smartphone microphones, Acoustic & 5 cm & \xmark & \xmark & \xmark\;Train, \checkmark\;ML & \xmark & `a-z' & \LEFTcircle & \LEFTcircle & 97.7\% \\ \hline

KeystrokeSniffer \cite{huang2024keystrokesniffer} & Smartphone microphones, Acoustic & 15 cm & \xmark & \xmark & \checkmark\;Train, \checkmark\;ML & \checkmark & `a-z' + 6 Special keys & \LEFTcircle & \CIRCLE & 79.5\% \\ \hline

Auditory Eyesight \cite{tu2023auditory} & Smartphone microphones, Acoustic & 50 cm & \xmark & \xmark & \xmark\;Train, \xmark\;ML & \xmark & 54 Commonly used keys & \LEFTcircle & \LEFTcircle & 90.8\% \\ \hline

UbiK \cite{wang2014ubiquitous} & Smartphone microphones, Acoustic & 20 cm & \xmark & \xmark & \checkmark\;Train, \checkmark\;ML & \xmark & 56 Commonly used keys & \LEFTcircle & \LEFTcircle & 95.0\% \\ \hline

Giallanza \textit{et al}. \cite{giallanza2019keyboard} & Smartphone microphones, Acoustic & 5 cm & \xmark & \xmark & \checkmark\;Train, \checkmark\;ML & \xmark & `a-z' + 8 Special keys & \LEFTcircle & \Circle & 70.6\% \\ \hline

SIA \cite{meteriz2021sia} & Smartwatch microphone, Acoustic & 5 cm & \xmark & \xmark & \checkmark\;Train, \checkmark\;ML & \xmark & `a-z' + `0-9' & \LEFTcircle & \LEFTcircle & 85\% \\ \hline

Bai \textit{et al}. \cite{bai2021know} & Smartphone microphones, Acoustic & 5 cm & \xmark & \xmark & \checkmark\;Train, \checkmark\;ML & \checkmark & `a-z' + 6 Special keys & \LEFTcircle & \LEFTcircle & 91.5\% \\ \hline

Slater \textit{et al}. \cite{slater2019robust} & Single microphone, Acoustic & \textless\;1 m & \xmark & \xmark & \checkmark\;Train, \checkmark\;ML & \xmark & Entire keyboard (68 keys) & \LEFTcircle & \Circle & 84.6\% \\ \hline

Cecconello \textit{et al}. \cite{cecconello2019skype} & Single microphone, Acoustic & 5 cm & \xmark & \xmark & \checkmark\;Train, \checkmark\;ML & \xmark & `a-z' & \LEFTcircle & \LEFTcircle & 91.7\% \\ \hline

S\&T \cite{compagno2017don} & Single microphone, Acoustic & 5 cm & \xmark & \xmark & \checkmark\;Train, \checkmark\;ML & \xmark & `a-z' & \LEFTcircle & \LEFTcircle & 83.2\% \\ \hline

Zhuang \textit{et al}. \cite{zhuang2009keyboard} & Single microphone, Acoustic & - & \xmark & \xmark & \checkmark\;Train, \checkmark\;ML & \xmark & `a-z' + 4 Special key & \Circle & \Circle & 92.0\% \\ \hline

Berger \textit{et al}. \cite{berger2006dictionary} & Single microphone, Acoustic & - & \xmark & \xmark & \xmark\;Train, \xmark\;ML & \checkmark & `a-z' & \LEFTcircle & \LEFTcircle & 90.0\% \\ \hline

Asonov \textit{et al}. \cite{asonov2004keyboard} & Single microphone, Acoustic & 1 m & \checkmark & \xmark & \checkmark\;Train, \checkmark\;ML & \checkmark & `a-z' + 4 Special key & \LEFTcircle & \Circle & 79.0\% \\ \hline

OverHear \cite{wijewickrama2023overhear} & Mic \& Accelerometers, Acoustic + Motion & 5 cm & \xmark & \xmark & \checkmark\;Train, \checkmark\;ML & \checkmark & `a-z' & \LEFTcircle & \LEFTcircle & 77.0\% \\ \hline

Fang \textit{et al}. \cite{fang2018no} & 1Tx \& 1Rx, WiFi CSI & 50 cm & \xmark & \xmark & \xmark\;Train, \xmark\;ML & \xmark & `a-z' & \Circle & \LEFTcircle & 95.3\% \\ \hline

WiKey \cite{ali2015keystroke} & 2 Tx ants \& 3 Rx ants, WiFi CSI & 30 cm & \xmark & \xmark & \checkmark\;Train, \checkmark\;ML & \xmark & `a-z' + `0-9' + 1 Special key & \Circle & \Circle & 96.4\% \\ \hline

Chen \textit{et al}. \cite{chen2015tracking} & SDR + FPGAs + 5 Rx ants, 2.4GHz RF signal & 5 m & \checkmark & \xmark & \checkmark\;Train, \checkmark\;ML & \checkmark & `a-z' + 1 Special key & \Circle & \Circle & 91.8\% \\ \hline

Marquardt \textit{et al}. \cite{marquardt2011sp} & Accelerometer & 5 cm & \xmark & \xmark & \checkmark\;Train, \checkmark\;ML & \checkmark & `a-z' & \Circle & \Circle & 80.0\% \\ \hline
\rowcolor{lightblue}
\textbf{This work} & \textbf{Piezo-based RFID tag, Acoustic} & \textbf{8 m} & \textbf{\checkmark} & \textbf{\checkmark} & \textbf{\checkmark\;Train, \checkmark\;ML} & \textbf{\xmark}  & \textbf{Entire keyboard} & \CIRCLE  & \CIRCLE  & \textbf{98.1\%} \\

\hline
\end{tabular}
}
\label{tab:related_work}
\end{table*}

%% file: 9_conclusion.tex
In this paper, we presented \pname, an RF backscatter system composed of a compact backscatter tag and an RF reader for keystroke eavesdropping attacks. The tag features two magnetically-coupled LC resonators, allowing it to convert keystroke-induced emanations into the frequency shifts of its backscatter signal.
It also achieves spectral separation between excitation and backscatter signals to mitigate self-interference.
The RF reader is equipped with a sophisticated signal processing pipeline that extracts reliable coarse-grained and fine-grained keystroke features for accurate typing recognition. Additionally, it leverages an LLM-guided online adaptation mechanism to enhance its generalizability.
Experimental results demonstrate the effectiveness of our attack in realistic settings. 
This work highlights both the potential and risks of combining RF sensing with LLMs for input inference, underscoring the need for future research on attack resilience and effective countermeasures.

%% file: 10_appendix.tex
\section*{Proof of Theorem~\ref{therem:fm}}
\label{app:theorem:fm}

As shown in Fig.~\ref{fig:tag_design}, this backscatter tag is composed of two LC resonators: 
a Parametric Enhancement Resonator (PER) 
and
a Voltage Sensing Resonator (VSR).
The PER is overlaid by VSR, and they are magnetically coupled.

\begin{figure}
    \centering
    \includegraphics[width=\linewidth]{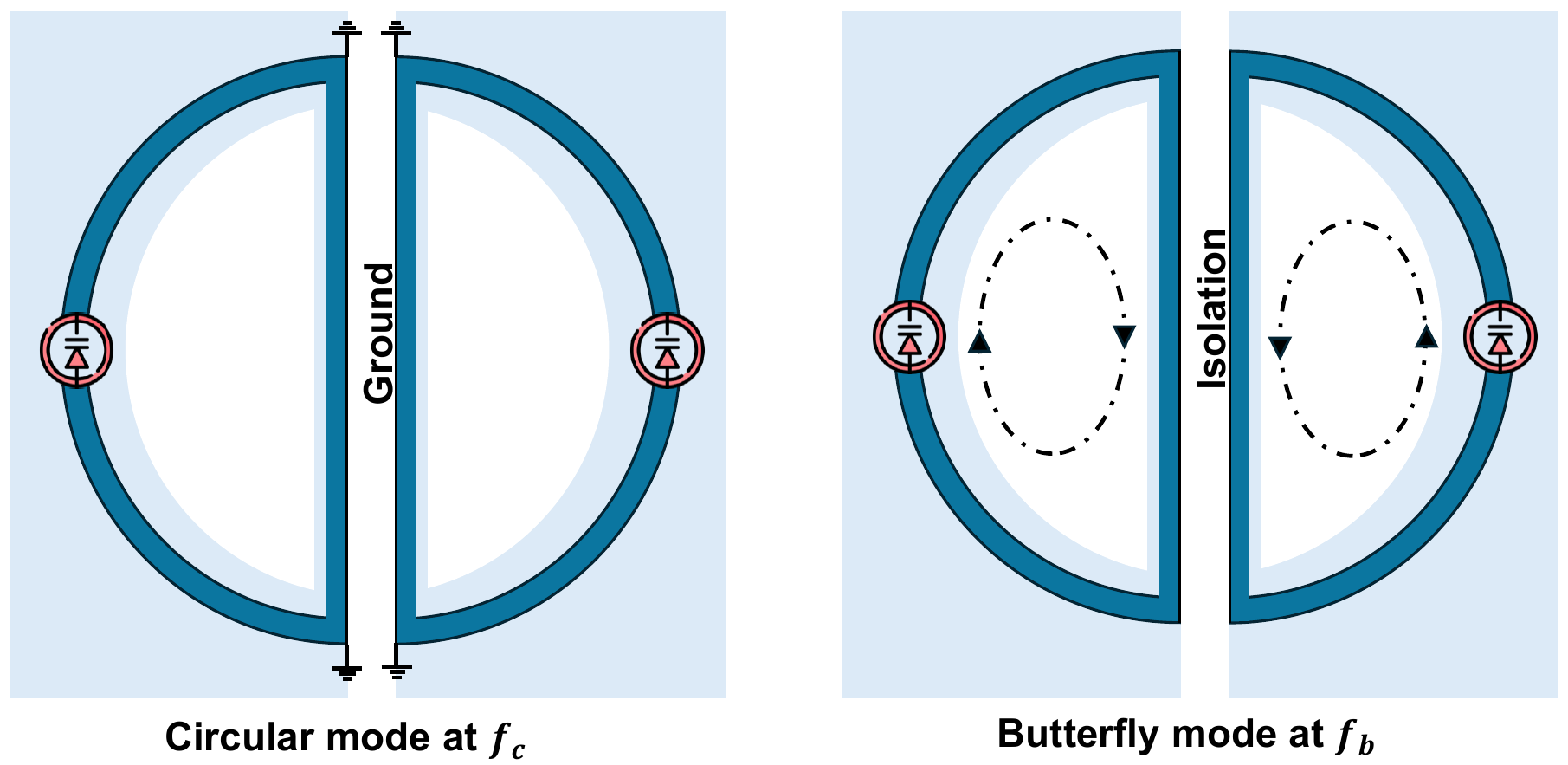}
    \caption{
The PER operates in the circular resonance mode (left), where the circuit has zero voltages in the center plane, making the circular mode equivalent to two half-loops sharing the same center ground. 
The PER operates in the butterfly resonance mode (right), where two separate current flows are confined within their individual meshes, making the butterfly mode equivalent to two half-loops that are electrically isolated in the center plane. 
}
\label{fig:PR}
\end{figure}

The PER consists of a circular-shaped conductor symmetrically divided by two gaps, each filled with a varactor (nonlinear capacitor) in a head-to-head configuration. This structure supports a circular resonance mode at frequency $f_{cr}$, where current flows around the ring. 
By connecting a horizontal conductor to the virtual voltage nodes of the circular mode, a second mode emerges at frequency $f_{br}$, enabling butterfly-shaped current flow in the corresponding circuit mesh. 
Fig.~\ref{fig:PR} illustrates these two modes of the PER. 
When the PER is excited by an external signal at the sum of the two resonance frequencies, i.e., $f_p \approx f_{cr} + f_{br}$, its nonlinear capacitance enables conversion of pumping power into enhanced backscattered signals near its resonant modes.

The VSR employs an 8-shaped conductor, with its ends connected by a bipolar junction transistor forming two head-to-head PN junctions. 
This implementation minimizes the number of solder joints, thereby decreasing parasitic resistance and increasing the quality factor. 
The base and emitter terminals are connected to a piezoelectric sensor, such that the VSR's resonance frequency can be modulated by pressure-induced voltages. Additionally, the 8-shaped layout helps suppress environmental electromagnetic artifacts.

When a bias voltage $v_s$ is applied to the sensing electrodes, it alters the varactor capacitance, thereby shifting the resonance frequency $f_1$ as follows:
\begin{equation}
\label{eq:f1}
\begin{aligned}
f_{1}
&= \frac{1}{2\pi} \sqrt{\frac{2}{L_1 C_1}} 
= \frac{1}{2\pi} \sqrt{\frac{2}{L_1 C_{10}(1 - v_s/\phi_1)^{-\lambda_1}}} \\
&\approx \frac{1}{2\pi} \sqrt{\frac{2}{L_1 C_{10}}} \left(1 - \frac{\lambda_1 v_s}{2\phi_1}\right)
\equiv f_{10}\left(1 - \frac{\lambda_1 v_s}{2\phi_1}\right),
\end{aligned}
\end{equation}
where $f_{10}$ denotes the zero-bias resonance frequency. Eqn~\eqref{eq:f1} shows that when the sensing voltage $v_s$ is much smaller than the junction potential $\phi_1$, the resonance frequency $f_1$ is approximately linearly related to $v_s$. 
This standalone resonator can be characterized using loop antennas connected to a network analyzer; however, this direct approach suffers from limited sensitivity, especially when the resonator is located far from the measurement antennas.

To enhance the remote detectability of the VSR, it is physically overlaid onto the perimeter of the PER, which functions as a local signal enhancer. To understand the coupling mechanism between these resonators, we analyze the relation between the electromotive force $\xi$ and the induced current $I$ in each resonator:
\begin{equation}
\label{eq:xi1}
\begin{aligned}
\xi_1 &= I_1 Z_1 + j 2\pi f M I_2 \\
      &= I_1 \left(R_1 + j 2\pi f L_1 - \frac{j}{2\pi f C_1}\right) 
      + j 2\pi f M I_2,
\end{aligned}
\end{equation}
\begin{equation}
\label{eq:xi2}
\begin{aligned}
\xi_2 &= I_2 Z_2 + j 2\pi f M I_1 \\
      &= I_2 \left(R_2 + j 2\pi f L_2 - \frac{j}{2\pi f C_2}\right) 
      + j 2\pi f M I_1,
\end{aligned}
\end{equation}
where $\xi_1$ and $\xi_2$ are the electromotive force for VSR and PER, respectively; $I_1$ and $I_2$ are their current. 
$R$, $L$, and $C$ represent their effective resistance, inductance, and capacitance, respectively; $M$ is the mutual inductance.

Solving for $I_2$ yields the induced current in the PER:

\begin{equation}
\label{eq:I2}
\begin{aligned}
I_2 &= \frac{Z_1 \xi_2 - j M 2\pi f\, \xi_1}{Z_1 Z_2 + M^2 (2\pi f)^2} \\
&= \frac{ 
\left( R_1 + j 2\pi f L_1 - j \frac{(2\pi f_1)^2 L_1}{2\pi f} \right) \xi_2 
- j 2\pi f M \xi_1
}{\left[
\begin{aligned}
& R_1 R_2 + (2\pi f)^2 M^2 
- \dfrac{L_1 L_2 (f^2 - f_1^2)(f^2 - f_2^2)}{f^2} \\
& + \dfrac{j}{2\pi f} 
\left(
L_1 R_2 \left((2\pi f)^2 - (2\pi f_1)^2\right) \right. \\
& \quad\quad 
\left. + L_2 R_1 \left((2\pi f)^2 - (2\pi f_2)^2\right) 
\right),
\end{aligned}
\right]},
\end{aligned}
\end{equation}
where $f_1$ and $f_2$ denote the stand-alone resonance frequencies of the VSR and PER, respectively.

When the coupled-mode frequency significantly deviates from the isolated-mode frequencies, both the term $R_1$ in the numerator and $R_1 R_2$ in the denominator can be neglected. 
To ensure $I_2$ is real-valued at resonance, the following condition must hold:
\begin{equation}
\label{eq:resonance_condition}
M^2 f^2 - L_1 L_2 \frac{(f^2 - f_1^2)(f^2 - f_2^2)}{f^2} = 0.
\end{equation}

Solving Eqn~\eqref{eq:resonance_condition} yields two coupled-mode frequencies:
\begin{equation}
\label{eq:fL}
f_L^2 = \frac{2 f_2^2}{f_2^2/f_1^2 + 1 + \sqrt{\left(f_2^2/f_1^2 - 1\right)^2 + 4\kappa^2 f_2^2/f_1^2}},
\end{equation}

\begin{equation}
\label{eq:fH}
f_H^2 = \frac{2 f_2^2}{f_2^2/f_1^2 + 1 - \sqrt{\left(f_2^2/f_1^2 - 1\right)^2 + 4\kappa^2 f_2^2/f_1^2}},
\end{equation}
where $\kappa^2 = M^2/(L_1 L_2)$ characterizes the coupling strength.

Defining $u_L = f_2^2/f_L^2$ and $u = f_2^2/f_1^2$, we can rewrite Eqn~\eqref{eq:fL} as:
\begin{equation}
\label{eq:rewrite}
u + 1 + \sqrt{(u - 1)^2 + 4\kappa^2 u} = 2u_L.
\end{equation}

Taking the derivative of both sides of Eqn~\eqref{eq:rewrite} with respect to $u$, we have 
\begin{equation}
\label{eq:derivative1}
1 + \frac{(u - 1) + 2\kappa^2}{\sqrt{(u - 1)^2 + 4\kappa^2 u}} = \frac{2 du_L}{du} =\frac{2 f_1^3 \, df_L}{f_L^3 \, df_1},
\end{equation}
where the final equality follows from the chain rule:
$du = -\frac{2 f_2^2}{f_1^3} df_1$ and
$du_L = -\frac{2 f_2^2}{f_L^3} df_L$.
%
%
Eqn~\eqref{eq:derivative1} can be rearranged to be: 
\begin{align}
\label{eq:dfL_df1}
\frac{df_L}{df_1} &= \frac{f_L^3}{2f_1^3} \left( 1 + \frac{(u - 1) + 2\kappa^2}{\sqrt{(u - 1)^2 + 4\kappa^2 u}} \right) \notag \\
&= \left( \frac{f_L^3}{2f_1^3} \right)
\left( 1 + 
\frac{ \frac{f_2^2}{f_1^2} - 1 + 2 \left( \frac{f_1^2}{f_L^2} - 1 \right)\left( \frac{f_2^2}{f_L^2} - 1 \right) }
{ \left| 2\frac{f_2^2}{f_L^2} - \frac{f_2^2}{f_1^2} - 1 \right| }
\right),
\end{align}
where the second equation holds because the coupling efficiency can be solved from Eqn~\eqref{eq:fL} to be 
$\kappa^2 = \big( \frac{f_1^2}{f_L^2} - 1 \big) \big( \frac{f_2^2}{f_L^2} - 1 \big)
$.


Eqn~\eqref{eq:dfL_df1} captures how the lower resonance frequency $f_L$ of the coupled system varies with the standalone frequency $f_1$ of the VSR. 
This modulation ratio is also affected by the stand-alone resonance frequency $f_2$ of the PER as well as their coupled resonance frequency $f_L$.

When the pump frequency $f_p$ deviates slightly from $f_{cr} + f_{br}$, signal regeneration occurs most efficiently when the reactance-to-resistance ratios match in both circular and butterfly resonance paths,
i.e.,
\begin{equation}
\label{eq:regeneration}
\frac{X_c}{R_c} = \frac{2(f_c - f_{cr}) 2\pi L_d}{ R_c} = \frac{X_b}{R_b} = \frac{2(f_b - f_{br}) 2\pi L_b}{R_b},
\end{equation}
where $f_c$ is the actual oscillation frequency, which is slightly different from its resonance frequency $f_{cr}$. 
Substituting $f_p = f_c + f_b$, we have
\begin{equation}
\label{eq:fc}
f_c = \frac{f_{cr} L_c / R_c - f_{br} L_b / R_b + f_p L_b / R_b}{L_c / R_c + L_b / R_b}.
\end{equation}

Since the VSR modulates $f_{cr}$ without affecting $f_{br}$, we have $f_{cr} = f_L$, where $F_L$ is defined in Eqn~\eqref{eq:fL}.
Since the voltage sensing resonator can somehow modulate the circular mode resonance frequency $f_{cr}$ without affecting the butterfly mode resonance frequency $f_{br}$, the value of $f_c$ in Eqn~\eqref{eq:fc} can also be effectively modulated. The oscillation frequency shift $\partial f_c$ and the resonance frequency shift $\partial f_L$  are correlated by the partial derivative relation obtained from Eqn~\eqref{eq:fc}, 
i.e.,
\begin{equation}
\label{eq:dfc_dfl}
\frac{\partial f_c}{\partial f_L} = \frac{L_c / R_c}{L_c / R_c + L_b / R_b}.
\end{equation}

By plugging the derivative of Eqn~\eqref{eq:f1} and 
plugging this derivative along with 
Eqn~\eqref{eq:dfL_df1} into Eqn~\eqref{eq:dfc_dfl}, we obtain the complete derivative with respect to $v_s$:

\begin{equation}
\label{eq:dfc_dvs}
\begin{aligned}
\frac{\partial f_c}{\partial v_s} 
&= \frac{\partial f_c}{\partial f_L} 
\cdot \frac{\partial f_L}{\partial f_1} 
\cdot \frac{\partial f_1}{\partial v_s} \\
&= \left( \frac{L_c / R_c}{L_c / R_c + L_b / R_b} \right)
\cdot \left( \frac{f_L^3}{2 f_1^3} \right) \\
&\quad \cdot \left[ 1 + 
\frac{ f_2^2 / f_1^2 - 1 }
{ \left| 2 f_2^2 / f_L^2 - f_2^2 / f_1^2 - 1 \right| } \right. \\
&\quad\quad \left.
+ \frac{2 (f_1^2 / f_L^2 - 1)(f_2^2 / f_L^2 - 1) }
{ \left| 2 f_2^2 / f_L^2 - f_2^2 / f_1^2 - 1 \right| } \right]
\cdot \left( - \frac{\lambda_1 f_{10}}{2 \varphi_1} \right).
\end{aligned}
\end{equation}

Eqn~\eqref{eq:dfc_dvs} reveals a closed-form expression for how the oscillation frequency $f_c$ is linearly modulated by the piezoelectric bias voltage $v_s$.
This completes the proof.